# Quantifying the Influence of Combined Lung and Kidney Support Using a Cardiovascular Model and Sensitivity Analysis-Informed Parameter Identification

Jan-Niklas Thiel[a,*], Ana Martins Costa[b], Bettina Wiegmann[c,d,e], Jutta Arens[b], Ulrich Steinseifer[a], Michael Neidlin[a]

**Affiliation:**
a) Medical Faculty, Department of Cardiovascular Engineering, Institute of Applied Medical Engineering, RWTH Aachen University, Forkenbeckstr. 55, 52074 Aachen, Germany
b) Engineering Organ Support Technologies group, Department of Biomechanical Engineering, University of Twente, Drienerlolaan 5, 7522 NB, Enschede, the Netherlands
c) Department for Cardiothoracic, Transplantation and Vascular Surgery, Hannover Medical School, Carl-Neuberg-Straße 1, 30625 Hannover, Germany
d) Implant Research and Development (NIFE), Lower Saxony Center for Biomedical Engineering, Stadtfelddamm 34, 30625 Hannover, Germany
e) German Center for Lung Research (DZL), Carl-Neuberg-Straße 1, 30625, Hannover, Germany

**\*Correspondence:**
Name: Jan-Niklas Thiel
Address: Institute of Applied Medical Engineering, Forkenbeckstr. 55
52074 Aachen, Germany
Email address: thiel@ame.rwth-aachen.de

**Funding:**
Funded by the Deutsche Forschungsgemeinschaft (DFG, German Research Foundation) SPP2014 "Towards the Artificial Lung" – project number: 447746988.

**Conflict of interest:**
All the authors have nothing to disclose.

**Authors' contributions:**
**Jan-Niklas Thiel:** Conceptualization, Data curation, Formal analysis, Investigation, Methodology, Software, Validation, Visualization, Writing –original draft. **Ana M. Costa:** Writing – review & editing. **Bettina Wiegmann:** Funding acquisition, Project administration, Resources, Data Curation, Writing – review & editing. **Jutta Arens:** Funding acquisition, Project administration, Writing – review & editing. **Ulrich Steinseifer:** Funding acquisition, Project administration, Resources, Supervision, Writing – review & editing. **Michael Neidlin:** Conceptualization, Funding acquisition, Project administration, Supervision, Writing – review & editing.

**Abstract**

Combined extracorporeal membrane oxygenation (ECMO) and continuous renal replacement therapy (CRRT) pose complex hemodynamic challenges in intensive care. In this study, a comprehensive lumped parameter model (LPM) is developed to simulate the cardiovascular system, incorporating ECMO and CRRT circuit dynamics. The model is used to analyze nine CRRT-ECMO connection schemes under varying flow conditions. Using a robust parameter identification framework based on global sensitivity analysis (GSA) and multi-start gradient-based optimization, we calibrated the model on 30 clinical data points from eight veno-arterial ECMO patients.

Our results indicate that CRRT has a significant impact on the cardiovascular system, with changes in pulmonary artery pressure of up to 202.5 %, highly dependent on ECMO flow. The GSA proved to be a powerful tool to improve the parameter estimation process. The established parameter estimation framework is fast and robust without the need for hyperparameter tuning and improves the parameter estimation process with an $R^2>0.98$ between simulation and experimental data. It uses modeling methods that could pave the way for real-time applications in intensive care.

This open-source framework provides a valuable tool for the systematic evaluation of combined ECMO and CRRT, which can be used to develop standardized treatment protocols and improve patient outcomes in critical care. In addition, as a digital twin, this model also provides a good basis for addressing research questions related to mechanical circulatory and respiratory support.

**Keywords:**
Lumped parameter modeling, cardiovascular modeling, global sensitivity analysis, parameter identification, extracorporeal membrane oxygenation, ECMO, continuous renal replacement therapy

# 1. Introduction

Extracorporeal membrane oxygenation (ECMO) is frequently used in intensive care medicine to treat cardiac and/or respiratory failure. It is used to partially or fully support heart/lung function in the sense of a "bridge to transplantation" or "bridge to recovery" [1]. An ECMO circuit consists of a pump, an oxygenator, and cannulae to drain and return oxygenated and decarboxylated blood. However, up to 70 % of patients develop an acute kidney injury during ECMO which can be attributed to the so-called lung-kidney crosstalk and the often reported fluid overload [2–5]. Additional continuous renal replacement therapy (CRRT) is therefore essential, but is also associated with higher morbidity and mortality [6–9].

Combined lung and kidney support can be realized in an integrated or separate manner. The separate approach requires separate vascular access for each device [7, 10–12]. This leads to greater circuit complexity with increased technical workload [13] and significantly larger artificial surface area, as well as circulatory complications such as bleeding [14], thrombus formation, and infection [15], resulting in significantly higher healthcare costs [16, 17]. In the integrated approach, CRRT is connected directly to the ECMO circuit. This presents challenges in controlling circuit pressure, which can lead to treatment interruptions, air entrapment, flow interruptions, and hemolysis [18, 19]. This combined therapy still lacks a gold standard and its connection configuration varies depending on the operator's practice and proficiency. In total, there are nine different possibilities for connecting the CRRT circuit to the ECMO circuit (depicted in Figure 2), which, together with the different cannula sizes, result in a wide variety of combinations.

In the context of the integrated approach, Xu et al. investigated the effect of different CRRT connection schemes on the pressures in the access and return lines of the CRRT circuit [20]. In-vitro experiments comparing six commonly used schemes showed significant pressure differences. In addition, a retrospective analysis of ten patients (seven veno-arterial ECMO (V-A ECMO), three veno-venous ECMO (V-V ECMO)) showed that changing the connection scheme can significantly reduce both access and return pressures. Wu et al. conducted a retrospective study with 100 patients who received combined ECMO and CRRT therapy [21]. Patients were divided into groups receiving separate and integrated support. The results showed that the separate group had a significantly longer CRRT initiation time and a shorter filter lifetime. In addition, local bleeding was only observed in the separate group in approximately 90 % of patients. While both studies highlight the advantages of the integrated approach for connecting CRRT to ECMO, the literature still lacks a systematic comparison of commonly used connection schemes under realistic patient conditions.

Lumped parameter models (LPM) offer a simplified approach to model the human cardiovascular system for analysis of ECMO hemodynamics. These models, analogous to electrical circuits, use ordinary differential equations (ODE) based on the conservation of mass and momentum [22]. Various LPMs exist, with the most prominent ones being those developed by Arts et al. and Shi et al., which address different aspects of cardiovascular dynamics [23, 24]. Models by Broomé et al. and Joyce et al. have investigated ECMO therapy, but neglect ECMO circuit dynamics, focusing instead on factors such as recirculation and gas exchange [25–31]. More sophisticated models, such as those by Fresiello et al. and Lazzari et al., integrate pump and cannula dynamics and serve as training tools, with a particular focus on usability through a graphical user interface rather than on the performance of the code [32, 33]. The majority of the models presented are not open-source or lack automated and robust parameter identification, which limits their usability for other research questions and their application in clinical decision support. The latter requires models that are based on patient-specific data and support real-time recalibration without the need for extensive user interaction or in-depth user knowledge, in line with the concept of digital twins defined by Viceconti et al. [34]. To date, there is no such model that includes both ECMO and CRRT therapy.

In contrast, for more general cardiovascular models, the topic of parameter identification has already been addressed in numerous studies. Local optimization algorithms, such as trust-region-reflective methods [35, 36] or Levenberg-Marquardt algorithms [37], are commonly used but rely heavily on initial values. In this context, the use of multi-start algorithms can provide guidance in finding a global optimum. The computation of gradients, which is typically performed by numerical differentiation, can result in inaccurate parameter estimation and numerical instability. Furthermore, this approach is known to be computationally expensive [38]. Automatic differentiation (AD) offers a faster and more stable alternative [38–42]. Global optimization techniques, such as the genetic algorithm, can also be utilized to identify a global optimum [43]. More sophisticated techniques, such as the unscented Kalman filter, can continuously update parameters in response to new measurements and are robust for dynamic, nonlinear systems [41, 44–47].

As cardiovascular models often consist of a high number of parameters, estimating their parameters often suffers from parameter identifiability and high computational cost. In this regard, parameter subset selection by ranking the influence of parameters by using global sensitivity analysis (GSA) and quantifying their independence of the effect on model outputs, also called parameter orthogonality, has been shown to improve the robustness and accuracy of parameter estimation [36–39, 41, 42, 48, 49].

Taken together, the combination of CRRT and ECMO therapy presents unanswered questions regarding optimal connection schemes and their impact on hemodynamics. LPMs appear to be a powerful tool for analyzing this problem, but challenges remain in achieving robust and systematic parameter identification. Overcoming these challenges is essential to improve the accuracy and applicability of these models in clinical decision support.

The aim of this study is to develop a cardiovascular model that includes both the detailed ECMO and CRRT system and allows arbitrary connection of both systems. A framework for fast and robust parameter identification based on multi-start gradient-based optimization will be linked to this model, which will be benchmarked on a number of V-A ECMO patients. Parameter selection will be based on Sobol indices from a GSA. Using this modeling framework, a comprehensive analysis of combined ECMO and CRRT therapy will be performed by investigating the impact of different CRRT connection schemes on patients and circuit dynamics at different ECMO flows.

## 2. Materials and Methods
### 2.1. Patient data

The clinical data set includes eight patients (five male, aged 36 – 61 years; three female, aged 32 – 76 years) treated with V-A ECMO, each with up to four measurements, resulting in a total of 30 data points. It includes 13 datapoints from male and 17 from female patients. These patients suffered from cardiogenic shock, cardiomyopathy, or received a heart or lung transplant. The measured hemodynamic parameters, along with their respective median, minimum and maximum values, are presented in Table 1.

**Table 1**
Measured hemodynamic parameters for 30 data points from eight V-A ECMO patients.

| Parameter | Median | (Min, Max) |
|---|---|---|
| Systolic pressure (SP) | 111.0 | (70.0, 183.0) mmHg |
| Diastolic pressure (DP) | 69.5 | (42.0, 88.0) mmHg |
| Mean arterial pressure (MAP) | 79.0 | (55.0, 113.0) mmHg |
| Mean pulmonary artery pressure (MPAP) | 19.5 | (6.0, 30.0) mmHg |
| Pulmonary capillary wedge pressure (PCWP) | 11.0 | (7.0, 15.0) mmHg |
| Cardiac output (CO) | 3.4 | (1.5, 7.5) L/min |
| Pump flow (PF) | 2.5 | (1.2, 3.8) mmHg |

### 2.2. Cardiovascular model

The model is inspired by Shi et al. [24] and consists of the heart, systemic and pulmonary circulation as shown in Figure 1. The heart is modeled as four chambers with variable elasticity [50], and the heart valves have a diodic pressure drop quadratically dependent on the flow to control the direction of blood flow. Both the systemic and pulmonary circulation are divided into aortic sinus, artery, arteriole, capillary, and venous segments. Large vessels such as the aortic sinus and artery are highly elastic, and their flow is inherently pulsatile. Therefore, they are modeled with resistance, compliance, and inertance (RCL) elements, where resistance [R] accounts for frictional losses, compliance [C] for elasticity, and inertance [L] for blood inertia. Smaller arterioles and capillaries are mainly dominated by resistance effects, while veins have the function of collecting and storing blood and therefore also have compliance effects.

In our model, circuits for mechanical circulatory support such as ECMO and left ventricular assist devices (LVAD) can be connected to any compartment of the patient's cardiovascular system, allowing different V-A and V-V ECMO configurations and different modes of LVAD support to be studied. Additionally, the model allows CRRT to be connected to the ECMO circuit in many different configurations. ECMO cannulas are modeled using a resistance based on data sheets of established Getinge products while taking compliance effects into account. The tubes are modeled with RCL elements similar to Lazzari et al. [33] and Fresiello et al. [32] and their resistance is approximated by the Hagen-Poiseuille law assuming a laminar flow regime.

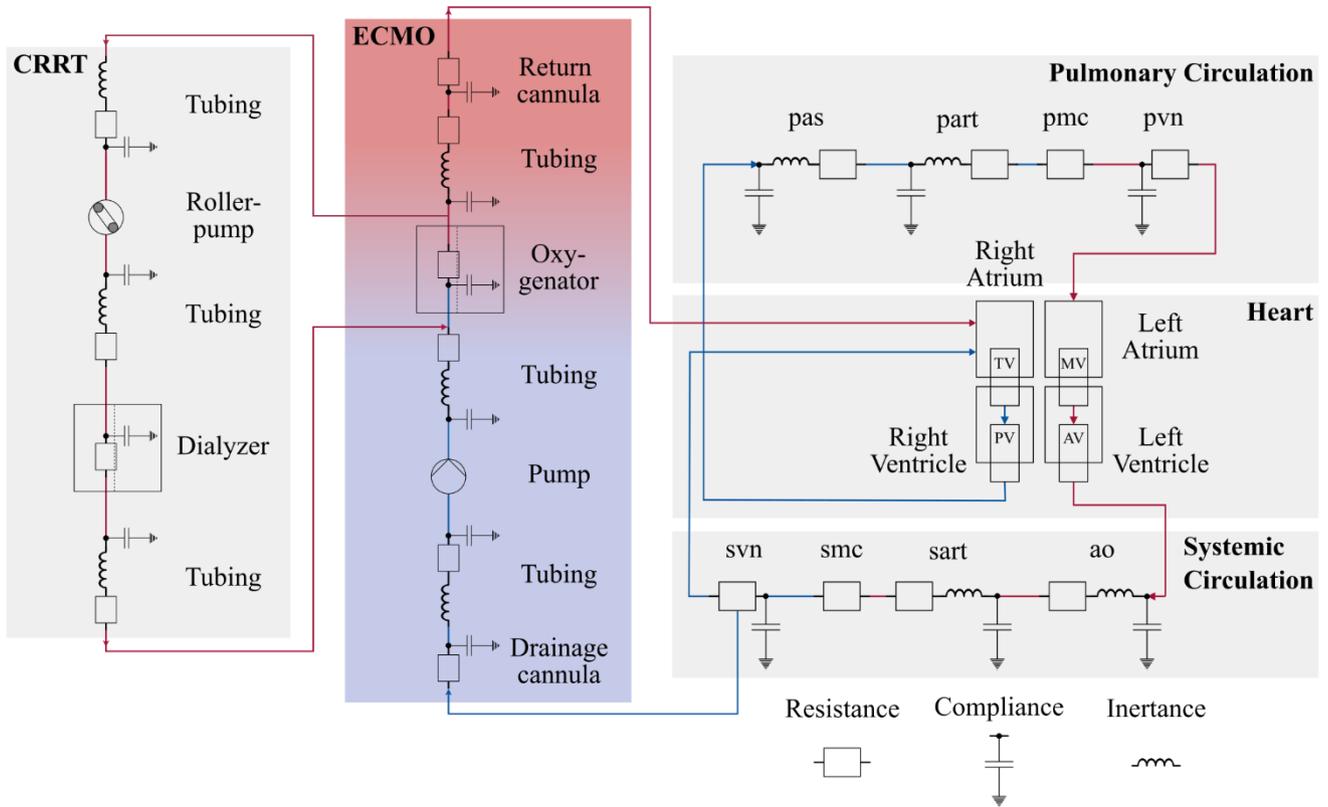

**Figure 1:** Overview of the lumped parameter model. Cardiovascular system inspired by Shi et al. [24]. pas: pulmonary aortic sinus, part: pulmonary artery, pmc: pulmonary microcirculation, pvn: pulmonary vein, ao: aorta, sart: systemic artery, smc: systemic microcirculation, svn: systemic vein, TV: tricuspid valve, PV: pulmonary valve, MV: mitral valve, AV: aortic valve, ECMO: extracorporeal membrane oxygenation, CRRT: continuous renal replacement therapy.

The ECMO pump is implemented using the analytical equations for rotary blood pumps (RBP) by Boes et al. [51]. This means that theoretically any RBP can be integrated into the LPM without having to change anything in the implementation. For this study, the parameters of this equation were adapted to the two blood pumps Rotaflow RF-32 (Getinge / Maquet Cardiopulmonary GmbH, Germany) and DP3 (Fresenius Medical Care AG, Germany). Roller pumps are modeled by specifying a constant flow rate. Both oxygenator and dialysis filter are modeled as RC elements, and their parameters were set to a constant flow rate. Both oxygenator and dialysis filter are modeled as RC elements, and their parameters were derived from the data sheets for Quadrox-i Adult and Small Adult (Getinge / Maquet Cardiopulmonary GmbH, Germany), Nautilus MC3 (MC3 Cardiopulmonary, USA) and Prismaflex M180 (Baxter International Inc., USA), assuming a linear relationship between pressure drop and flow.

The model compartments were initialized with physiological pressures representative of a healthy patient and a flow of zero. The model was implemented in *Python* using *JAX* with the possibility of just-in-time compilation and automatic differentiation. Therefore, all implemented functions are smooth, either inherently or by using techniques such as a smoothed Heaviside function. This ensures that the model is fully differentiable. The ODE system was solved using Dormand Prince's 8th order Runge-Kutta method, which is implemented in the *Diffrax* package. Adaptive time step sizing approach with an initial time step size of 0.0005 s was employed. The simulation time was set to 50 s to achieve cycle convergence. Only the last two cycles were evaluated.

We provide the full *Python* code with all described modules and parameter values at the GitHub link https://github.com/nikithiel/ECLIPSE. Parameter values used can also be found in Tables S-1, S-2, S-3 and S-4 in the appendix.

**2.3. Global sensitivity analysis**

The influence of the model parameters on the model outputs and their interactions between each other were quantified using the Sobol method [52], which is implemented in the *SALib* package. A perturbation of 25 % was applied to the initial values of the parameters and sampling was performed by Saltelli's extension of the Sobol' sequence. Since the results of a GSA are strongly dependent on the sample size, as clearly shown by Saxton et al. [53], a convergence analysis was performed, which can be found in the appendix in Figure S-1. A sample size of $N = 2^{11}$ was chosen with converging means and statistical deviations of less than 5 % from the mean. The influence of all $D = 24$ cardiovascular system (CVS) parameters was analyzed, resulting in $N(2D + 2) = 102400$ model evaluations. The results are systolic pressure [SP], diastolic pressure [DP], mean arterial pressure [MAP], mean pulmonary arterial pressure [MPAP], pulmonary capillary wedge pressure [PCWP], ECMO pump flow [PF] and cardiac output [CO]. The union of model parameters accounting for 90 % of the total sensitivity $S_T$ for each output was selected. This selection procedure was performed for initial parameter values representative of patients in normal, hypertensive, and hypotensive states to include the variability of GSA based on the initial parameter values.

### 2.4. Parameter identification

The selected parameters were used to minimize the sum of squared errors between the model predictions and the patient data described earlier. For this objective function, the parameter space was explored to verify that the function is smooth and differentiable within the relevant bounds. The results of this parameter space exploration can be found in Figure S-2 in the appendix. A gradient-based optimization algorithm with bound constraints was used, and the gradients were calculated using AD. We used *SciPy's* least squares optimizer with trust region reflective algorithm as it is particularly efficient to exploit the least squares structure of the objective function. Additionally, we used multi-start based on the Tiktak algorithm developed in Arnoud et al. [54] with 70 runs of local optimizations. This was all implemented using the *Estimagic* package.

### 2.5. Combined ECMO and CRRT study

For the analysis of combined lung and kidney support therapy, the clinically relevant CRRT connection schemes shown in Figure 2 were applied to the ECMO circuit. From the previously fitted parameters, a 69-year-old female patient with cardiogenic shock and acute right heart failure was selected, who also received renal support during her therapy. A constant flow through the CRRT circuit of 0.2 L/min was set.
In a first step, the influence of the different connection types on the entire cardiac cycle was investigated for a constant pump speed of 3425 rpm. Next, the influence of the different connection types was examined at speeds between 1000 and 5000 rpm, which corresponds to the typical operating points of the Rotaflow RF-32 pump. Here, we examined only cycle-averaged variables. Furthermore, a GSA was again used to systematically investigate the influence of these two factors. For this, resistances, compliances, and inertances of the CVS were grouped and used together with the locations of drainage and return cannula as discrete values and the pump speed of the ECMO circuit as inputs. We added 25 % perturbation to the continuous input parameters and used sampling as previously described. In this way, the effects of small changes in CVS parameters corresponding to clinical treatments of hypertension and hypotension (e.g. through inotropes or vasopressors) can be compared to changes in the extracorporeal system, such as ECMO pump speed and location of CRRT connection.

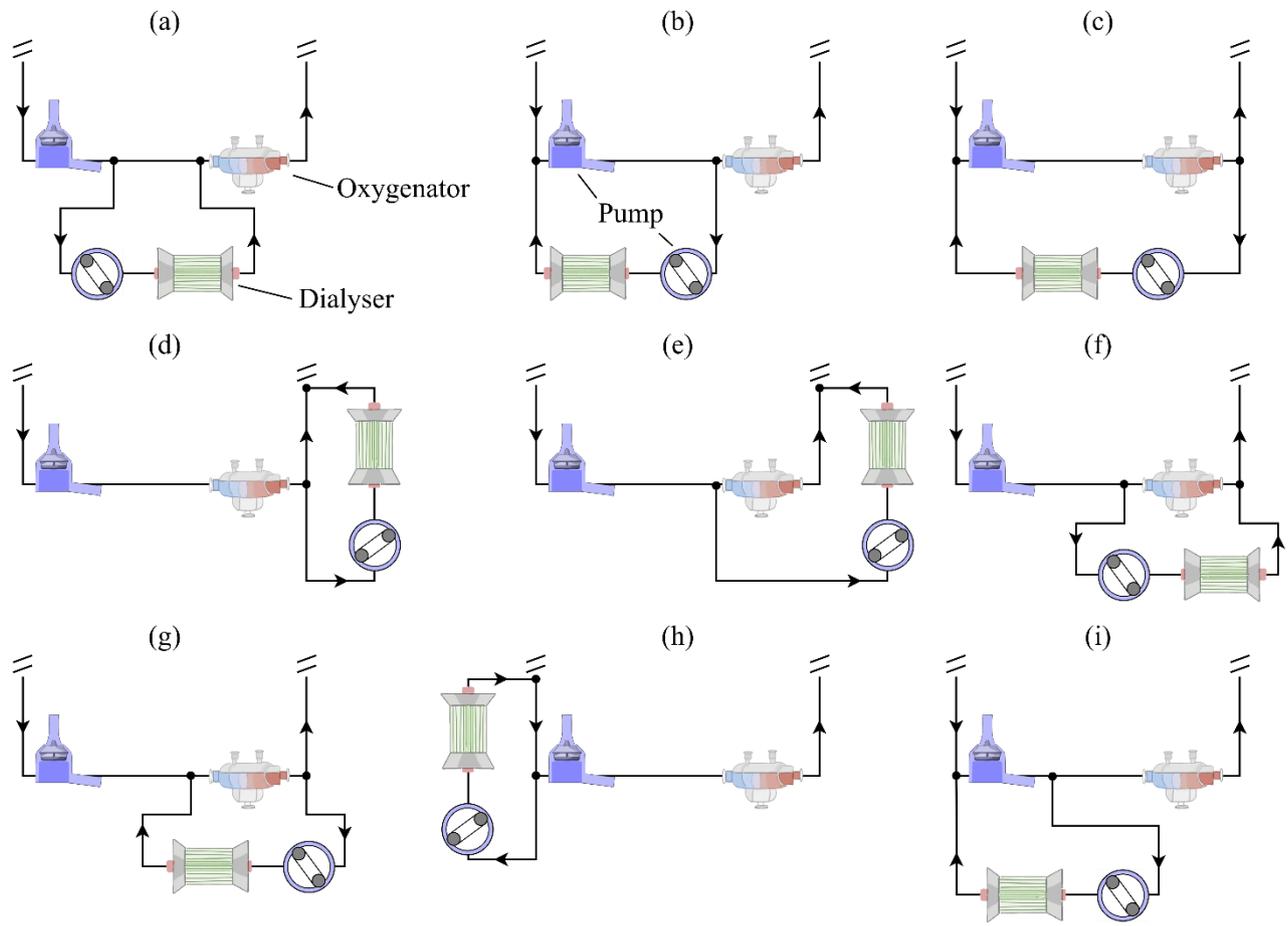

**Figure 2:** Variations of CRRT connections to the ECMO circuit in CRRT-flow direction: a) post pump – pre oxygenator (pre oxy), b) pre oxy – pre pump, c) post oxy – pre pump, d) post oxy – pre return cannula (pre RC), e) pre oxy – pre RC, f) pre oxy – post oxy, g) post oxy – pre oxy, h) pre pump – post drainage cannula (post DC), i) post pump – pre pump. Of note: position of tubing e.g. between pump and oxygenator (compare (b) and (i)) is also taken into account by making a distinction between post pump and pre oxy.

## 3. Results
### 3.1. Selection of most important model parameters for calibration

The total order sensitivity indices [$S_T$] for parameter values representative of a patient with normal systolic pressure are shown in Figure 3. This figure illustrates the effect of $S_T$ of cardiovascular model parameters on all clinical measurements. The difference between total order and first or sensitivity indices can be seen in the Figure S-3 of the appendix. This shows that all model parameters influence the model outputs in an independent way. From Figure 3 it can be seen that the systolic elastance of the left ventricle [Emaxlv] and the resistance of the systemic microcirculation [Rmc] have a significant influence on all outputs, as they account for 90 % of their total order sensitivity. The compliance of the systemic artery [Csart] mainly affects SP and DP. In contrast, the diastolic elastance of the left atrium [Edla] only affects PCWP. Both the systolic elastance of the right ventricle [Emaxrv] and the resistance of the systemic vein [Rsvn] mainly affect MPAP and PCWP. In contrast, the resistance of the pulmonary microcirculation [Rpmc] shows a high $S_T$ for MPAP.

Table 2 shows the sum of the effects of the most influential input parameters on all outputs. The following seven parameters, displayed in orange in Figure S-4 in the appendix, are selected for the parameter identification step: Rmc, Emaxlv, Rpmc, Emaxrv, Csart, Edla and Rsvn. The significant influence of Rmc and Emaxlv can also be observed here.

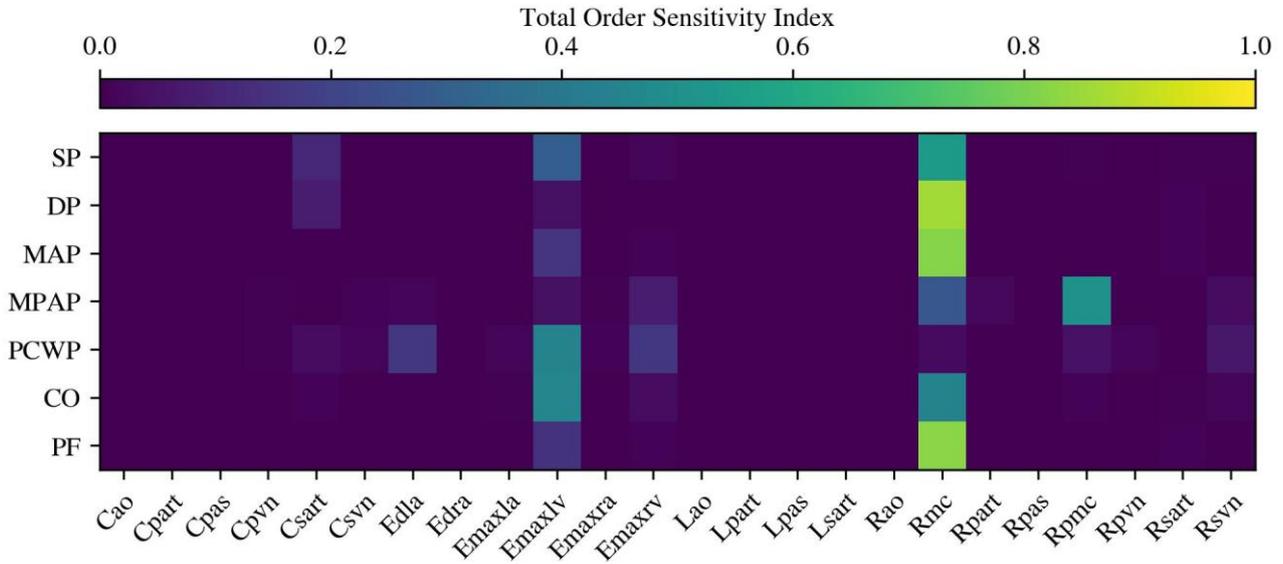

**Figure 3:** Total order sensitivity indices for patient with normal systolic pressure. Cao: compliance of aorta, Cpart: compliance of pulmonary artery, Cpas: compliance of pulmonary aortic sinus, Cpvn: compliance of pulmonary vein, Csart: compliance of systemic artery, Csvn: compliance of systemic vein, Edla: diastolic elastance of left atrium, Edra: diastolic elastance of right atrium, Emaxla: systolic elastance of left atrium, Emaxlv: systolic elastance of left ventricle, Emaxra: systolic elastance of right atrium, Emaxrv: systolic elastance of right ventricle, Lao: inertance of aorta, Lpart: inertance of pulmonary artery, Lpas: inertance of pulmonary aortic sinus, Lsart: inertance of systemic artery, Rao: resistance of aorta, Rmc: resistance of systemic microcirculation, Rpart: resistance of pulmonary artery, Rpas: resistance of pulmonary aortic sinus, Rpmc: resistance of pulmonary microcirculation, Rpvn: resistance of pulmonary vein, Rsart: resistance of systemic artery, Rsvn: resistance of systemic vein, SP: systolic pressure, DP: diastolic pressure, MAP: mean arterial pressure, MPAP: mean pulmonary artery pressure, PCWP: pulmonary capillary wedge pressure, CO: cardiac output, PF: pump flow.

**Table 2**
Cumulative total order sensitivity indices [$S_T$] for most influential model parameters for patient with normal systolic pressure. Rmc: resistance of the systemic microcirculation, Emaxlv: systolic elastance of the left ventricle, Rpmc: resistance of the pulmonary microcirculation, Emaxrv: systolic elastance of the right ventricle, Csart: compliance of the systemic artery, Edla: diastolic elastance of the left atrium, Rsvn: resistance of the systemic vein.

| Parameter | $\sum S_T$ |
|---|---|
| Rmc | 3.80 |
| Emaxlv | 1.60 |
| Rpmc | 0.60 |
| Emaxrv | 0.32 |
| Csart | 0.25 |
| Edla | 0.19 |
| Rsvn | 0.14 |

### 3.2. Model calibration

Figure 4 a) and b) illustrate a scatter plot of measured versus predicted pressures and flows for 30 data points of V-A ECMO patients. The corresponding boxplots are depicted in Figure 4 c) and d). The measured data cover a wide range of clinically relevant operating points of ECMO therapy, with pump flows between 1 and 4 L/min. These include patients with both hypotension (MAP < 65 mmHg) and hypertension (MAP > 92 mmHg), as well as patients with a normal MAP. A high-quality fit can be achieved with an $R^2$ of 0.99 for the pressures and 0.98 for the flows. It can be seen that the ECMO pump flows are overestimated and PCWP is underestimated.

### 3.3. Influence of CRRT connection scheme

This section examines the influence of different CRRT connection schemes at a constant pump speed of 3425 rpm, resulting in a flow of 3.7 L/min through the ECMO circuit. Figure 5 a) shows the arterial, pulmonary arterial and venous pressures for two cardiac cycles. It can be seen that the CRRT circuit has an impact on the patient's hemodynamics, with the greatest effect on the venous system, which has a median of 3.7 mmHg and an absolute variation of 1.0 mmHg (26.3 %), and the lowest effect on the arterial system, with a median of 70.3 mmHg and an absolute variation of 5.4 mmHg (7.7 %). The patient's blood pressure is lowest when the access line is downstream of the ECMO pump (post pump), and the return line is upstream of the oxygenator (pre oxy, for configuration see Figure 2a)). In contrast, the blood pressures are highest for the access line placed pre pump and the return line placed post drainage cannula (post DC) (configuration Figure 2h)). Figure 5 b) illustrates the right ventricular (RV) PV-loops for all connection schemes. The access line post pump and the return line pre oxygenator (configuration Figure 2a)) lead to a shift of the RV PV-loops to lower pressures and volumes and to a smaller stroke volume. For the combination pre pump – post DC which leads to the highest blood pressures in Figure 5 a), a reverse effect can be observed (configuration Figure 2h)), which leads to an absolute variation of approximately 27.7 mL (24.6 %) and 24.5 mL (24.0 %) for the end-diastolic and end-systolic volumes of the RV.

The pressures in the access and return lines of the CRRT circuit are illustrated in Figures 5 c) and 5 d). Both pressure locations show high variability, with a median of 161.4 and 258.2 mmHg, and an absolute variation of 296.6 mmHg (183.7 %) and 295.9 mmHg (114.6 %) for access and return line, respectively. All pressures are below the maximum and above the minimum pressure alarms that are typically set as standard for CRRT devices. Furthermore, all pressures are positive, except for the connection where the access and return lines are upstream of the ECMO pump (configuration Figure 2h)). Arterial, pulmonary arterial and venous pressures, as well as the pressure at the tip of the ECMO drainage cannula, are shown in Figure 6 a) and b) for different CRRT connection types and ECMO flows. Varying the ECMO pump speed between 1000 and 5000 rpm results in ECMO flows of up to 6 L/min. With increasing ECMO flow, arterial pressures increase for all combinations, whereas pulmonary arterial and venous pressures decrease. The influence of CRRT connection types increases with increasing speed. At an ECMO flow of 1.2 L/min, arterial and ECMO drainage pressures have a median of 49.4 and 1.7 mmHg, respectively, with an absolute variation of 1.4 mmHg (2.9 %) and 1.3 mmHg (79.8 %). However, at a significantly higher flow of 6 L/min, these values increase and decrease to 85.2 mmHg and -129.6 mmHg with an absolute variation of 10.5 mmHg (12.3 %) and 3.9 mmHg (3 %), respectively. At this maximum ECMO flow, different CRRT connection schemes can cause variations in pulmonary artery pressure of up to 5.3 mmHg (202.5 %).

It appears that the type of connection that provides the highest and lowest blood pressures depends on the ECMO flow.

The pressures in the access and return lines of the CRRT circuit are shown in Figure 6 c) and d). The same trend is seen as before. The mean pressures in the access and return lines of all combinations change drastically from 2.6 and 136.5 mmHg with an absolute variation of 81.3 mmHg (3163.8 %) and 81.6 mmHg (59.8 %) at 1.2 L/min to 397.4 and 459.5 mmHg with an absolute variation of 624.5 mmHg (157.1 %) and 622.9 mmHg (135.5 %) at 6 L/min, respectively. When the access line is connected downstream of the ECMO pump and the return upstream of the oxygenator (configuration Figure 2a)), pressures increase with increasing ECMO flow. This behavior is observed for most combinations. The connection pre pump – post DC (Figure 2h)) shows the opposite behavior. This decrease in pressure with increasing ECMO flow is similar for all combinations with return lines upstream of the ECMO pump (Figure 2 b), c), h), and i)). The maximum pressure alarms are reached at approximately 5 and 4 L/min for the access and return line, respectively. A minimum pressure alarm is only triggered for the return line at around 6 L/min.

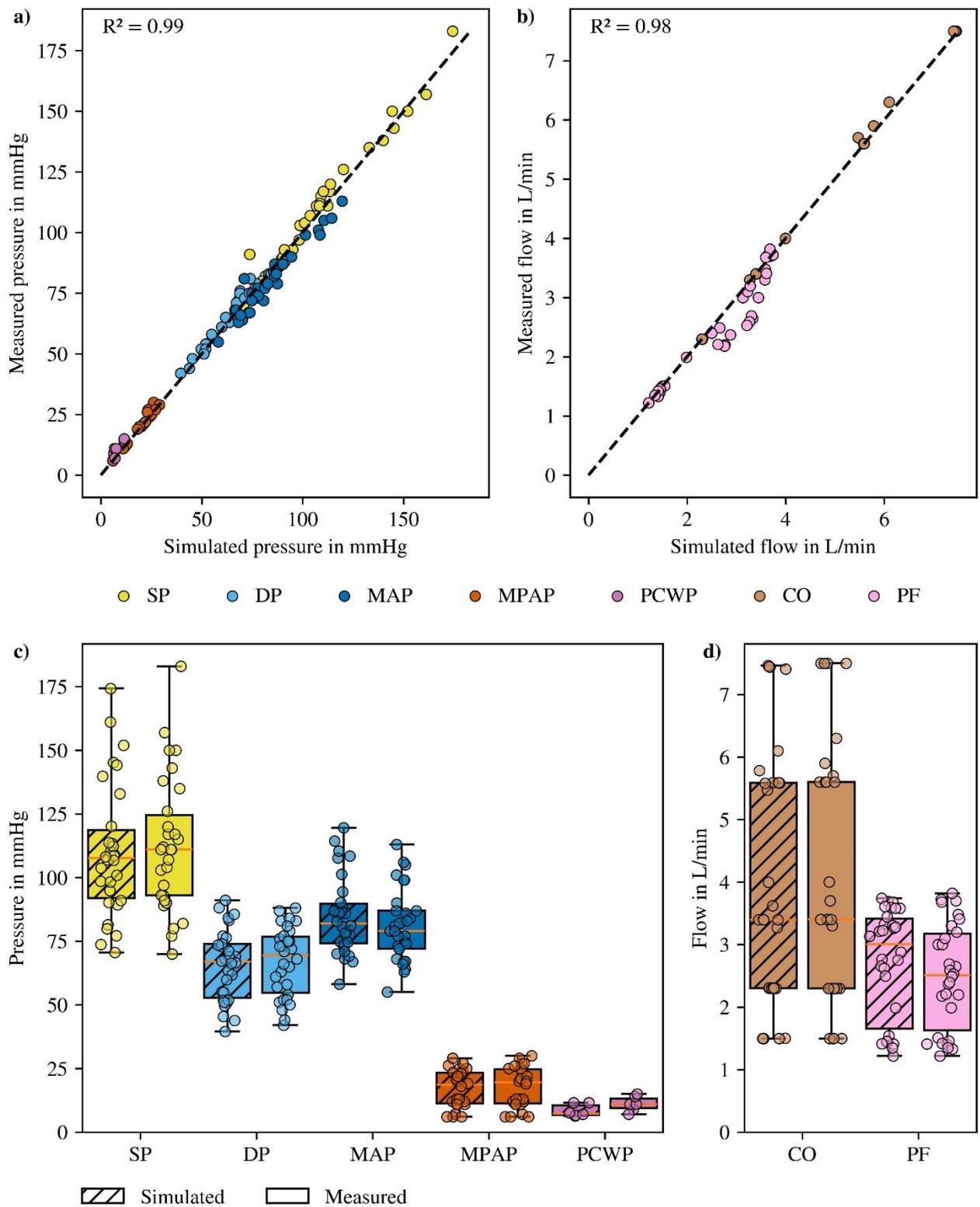

**Figure 4:** a) + b) Scatter plot of simulated data against measurement data for n = 30 datapoints of 8 V-A ECMO patients. c) + d) Box plots for simulated and measurement data for n = 30 datapoints of 8 V-A ECMO patients. SP: systolic pressure, DP: diastolic pressure, MAP: mean arterial pressure, MPAP: mean pulmonary artery pressure, PCWP: pulmonary capillary wedge pressure, CO: cardiac output, PF: pump flow.

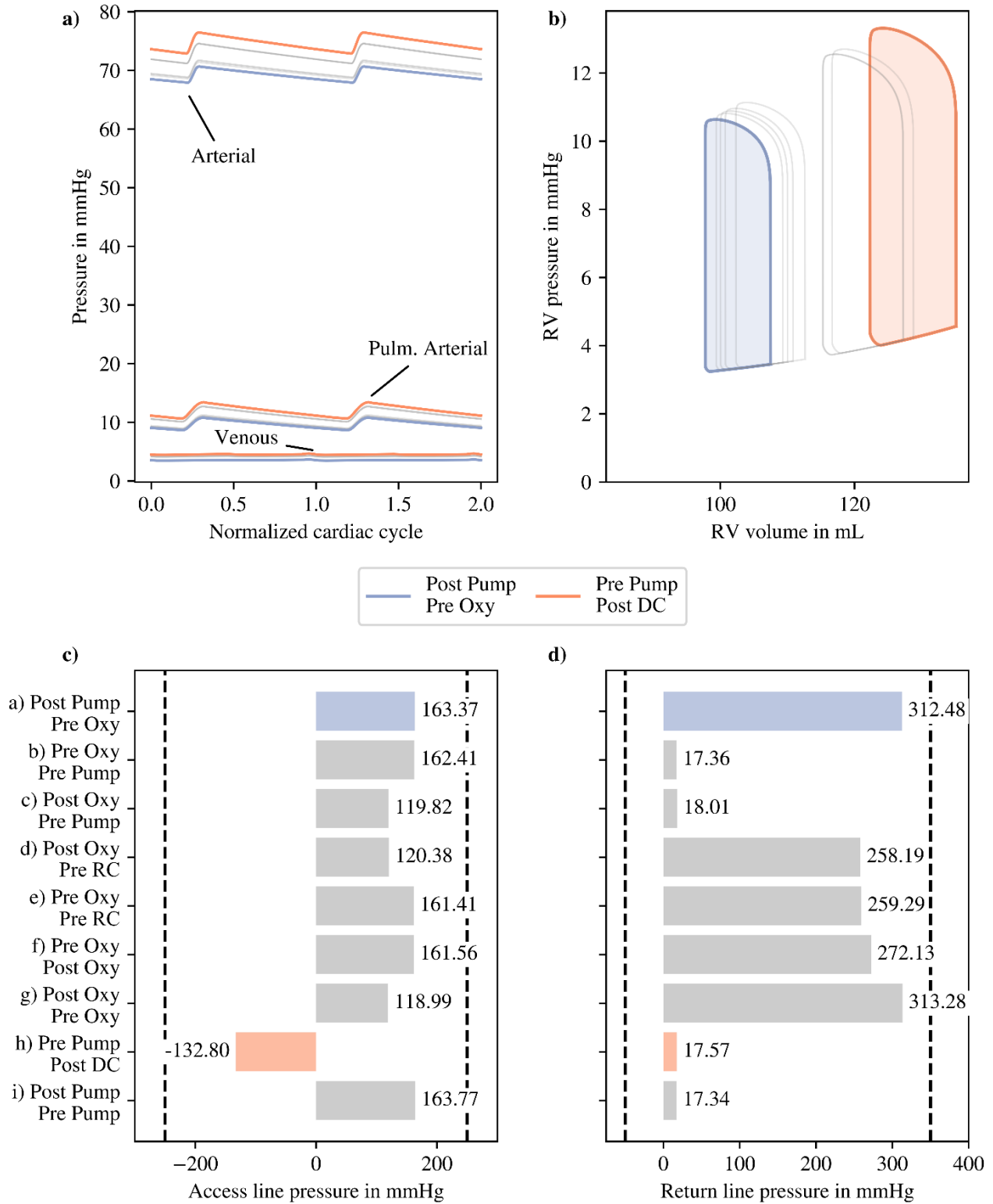

**Figure 5:** Influence of different CRRT connections schemes on a) cardiovascular system, b) right ventricular (RV) PV-loop and pressure of both access and return line of the CRRT circuit in c) and d), respectively. Common pressure alarms of CRRT circuit displayed in black dashed lines. Orange lines represent the configuration pre pump – post DC, blue lines post pump – pre oxy, grey lines the other configurations. DC: drainage cannula, RC: return cannula.

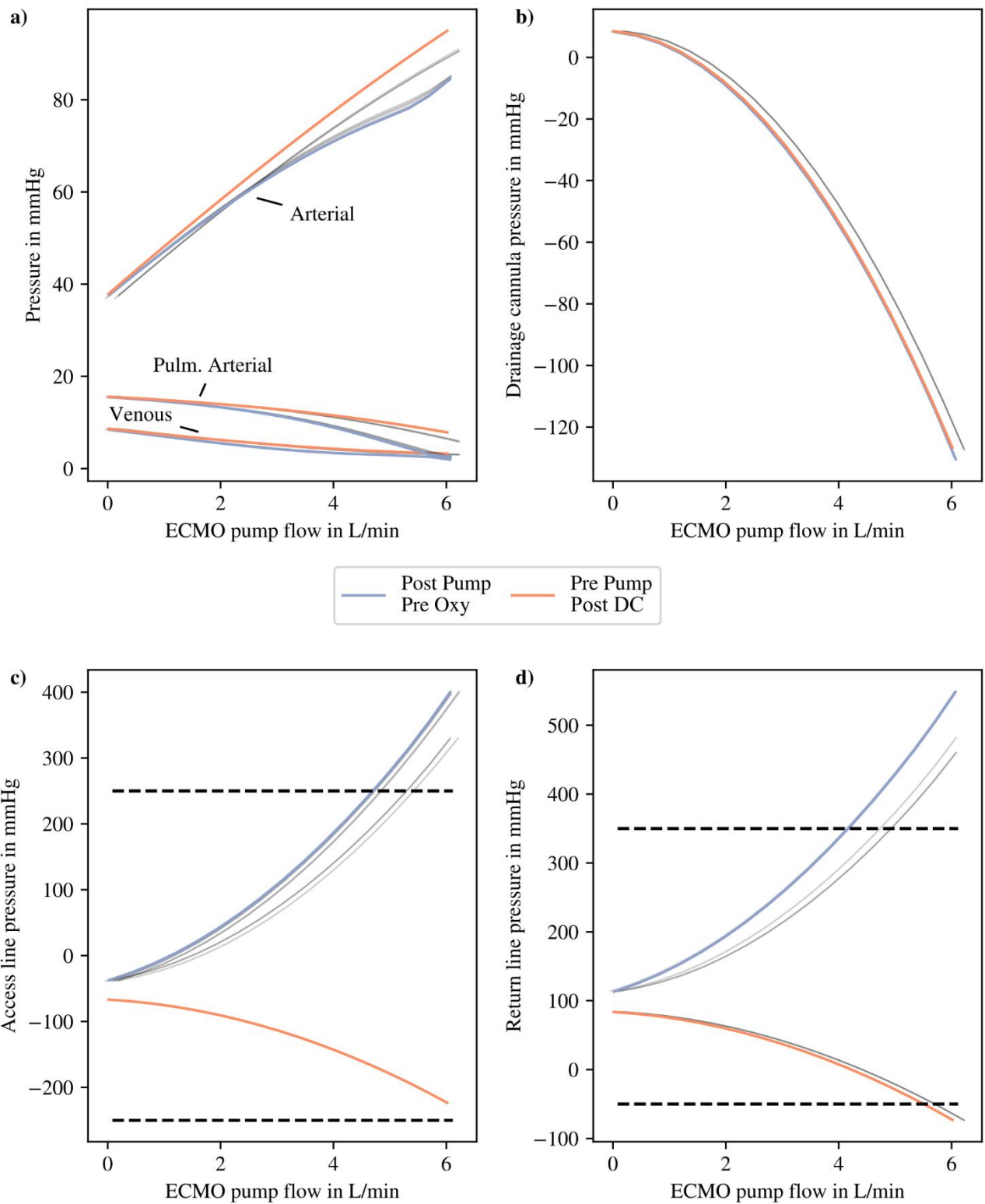

**Figure 6:** Influence of different CRRT connections schemes with increasing ECMO pump flow on a) cardiovascular system, b) pressure in ECMO drainage cannula (DC) and both access and return line of the CRRT circuit in c) and d), respectively. Common pressure alarms of CRRT circuit displayed in black dashed lines. Orange lines represent the configuration pre pump – post DC, blue lines post pump – pre oxy, grey lines the other configurations.
.

The total order sensitivity indices [$S_T$] for the grouped cardiovascular parameters R, C, E and L, for the position of the CRRT circuit access and return lines, and for the ECMO pump speed for all model outputs are shown in Figure 7. ECMO speed is the most sensitive parameter and R has a particularly large influence on the patient's hemodynamics. The position of both the access and return lines is as sensitive to MPAP and CO as C, E and L. This is supported by the cumulative $S_T$, which can be found in Figure S-5 in the appendix as well. In summary, both locations of the access and return lines contribute to 90 % of the sensitivity of MPAP and CO. As can be seen in Figure S-6 in the appendix, the GSA input samples resulted in a MAP between 50 and 100 mmHg and pump flows between 2 and 5 L/min, again representing normal, hypotensive and hypertensive states and the full range of ECMO operating conditions.

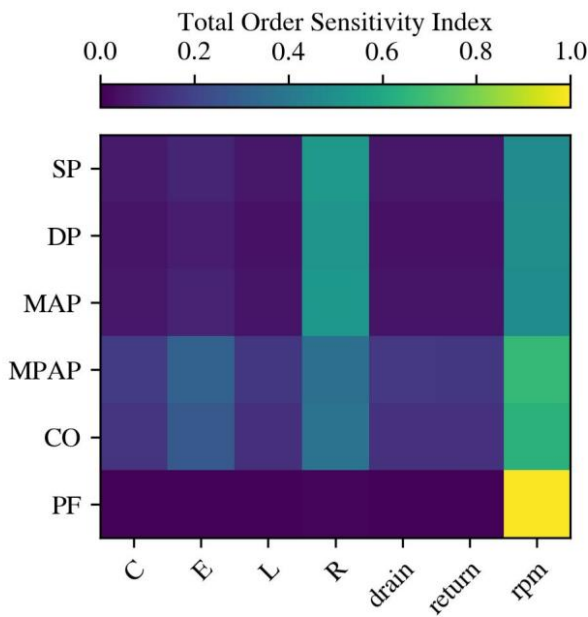

**Figure 7:** Total order sensitivity indices for V-A ECMO patient from Global sensitivity analysis including CRRT connection scheme and ECMO pump speed. Grouping of model parameters into resistances (R), compliances (C), and inertances (L) of the cardiovascular system and properties of both left and right ventricle (E).

## 4. Discussion

Combined ECMO and CRRT support is a complex therapy with two extracorporeal circuits that interact with both each other and the patient's cardiovascular system. There are no standardized guidelines for optimal connection schemes. In this study, we utilized a 0D computational model to investigate the mutual influence of CRRT connection, ECMO support and the patient's circulation. Additionally, we conducted a global sensitivity analysis and developed a robust parameter estimation pipeline using 30 data points from eight V-A ECMO patients. Our main findings are:

1. CRRT has a significant effect on the patient's cardiovascular system. This effect is highly dependent on the ECMO flow and leads to changes of up to 202.5 %.
2. GSA proves to be a powerful tool for enhancing the parameter estimation process.
3. The established parameter estimation framework is robust and does not require any hyperparameter tuning.

An ECMO flow of 3.7 L/min in our in-silico study at a constant pump speed of 3425 rpm is comparable to the retrospective study by Wu et al. [21], which reported flows of 3.1 and 4.4 L/min. Also, the pressures at the access and return lines of the CRRT circuit are consistent with this study and the study by Xu et al. [20] and vary greatly depending on the connection location in the ECMO circuit, as also reported by Kashani et al. [55]. Interestingly, while Xu et al. [20] observed negative pressures in the return line when connected upstream of the ECMO pump, both our results and those of Wu et al. [21] show positive return pressures for all configurations. This discrepancy could be due to the characteristics of the in-vitro setup used by Xu et al.

In method A from Wu et al., where the access line is connected downstream of the oxygenator and the return line upstream (our configuration Figure 2g)), the clinical data show a mean access line pressure of 175 mmHg (SD = 23 mmHg) and a mean return line pressure of 360 mmHg (SD = 8 mmHg). Our model predicts pressures of 119 and 313 mmHg, respectively. In method B, where the return line is changed to be upstream of the pump (our configuration Figure 2c)), the clinical data show a significantly lower mean return line pressure of 41 mmHg (SD = 13 mmHg) and an access line pressure of 171 mmHg (SD = 22 mmHg). Our model predicts 17 mmHg for the return line and 120 mmHg for the access line. These results confirm that our model can predict clinically realistic pressures in the CRRT circuit and shows the same trends when the return line position is changed.

The in-vitro data from Xu et al. [20] show that both the pressures in the access and return lines and the influence of the CRRT connection increase with ECMO flow, with the pressure in the access line triggering the maximum alarm at around 5.5 L/min. This observation is also reflected in the predictions of our model. Sansom et al. [56] have found that high access line pressures correlate with early CRRT circuit failure and suggest that these pressures should be kept below 190 mmHg. Our model suggests that this can be maintained for ECMO flows up to 4.7 L/min when the access line is connected downstream the oxygenator and the return line is placed upstream the oxygenator (Figure 2g)), the ECMO return cannula (Figure 2d)), or the ECMO pump (Figure 2c)). In contrast, connecting the access line after the ECMO pump and the return line before the oxygenator (Figure 2a)) results in a pressure higher than 190 mmHg at an ECMO flow of 4 L/min already. This insight can help to extend circuit life without increasing the cost and complexity of the extracorporeal circuit by adding pressure sensors, as suggested in Na et al. [57].

Pressures in the entire ECMO circuit are consistent with the values reported by Sidebotham et al. [58]. Wu et al. [21] recommend connecting the CRRT circuit downstream to upstream of the oxygenator as depicted in Figure 2g). Looking at our in-silico results, this seems to be a good choice. Using our model, we can additionally confirm the advice to connect the return line upstream of the ECMO pump for high ECMO flows, as this reduces the pressure in the return line and the risk of triggering internal alarms of the CRRT device. Although many authors report a high risk of air leakage in this configuration due to negative pressures in the ECMO circuit upstream of the pump [7, 11, 12, 19], Wu et al. did not observe this issue in their study. If the system setup does not allow connection of the return line upstream of the ECMO pump, or if pressure reduction is not critical, connection upstream of the ECMO return cannula as depicted in Figure 2d) provides a stable and safe alternative, as supported by the in-vivo data of Xu et al. [20]. This approach helps manage high pressures and reduces the burden on nursing staff, as described by Kashani et al. [11] and de Tymowski et al. [19], potentially leading to fewer CRRT interruptions and maintaining normal operation even at high ECMO flows. This is particularly important in patients with high fluid overload.

The results of the total order sensitivity index [$S_T$] indicate that the position of both the access and return lines have an influence on the hemodynamics of the patient, which is particularly significant for MPAP and CO. For most model parameters, the first order sensitivity index [$S_1$] for the different model parameters is negligibly small, indicating that their main effect is small. A closer look at the second order sensitivity indices [$S_2$], which describe pairwise interactions, reveals strong interactions between access and return line locations and other parameters. The strongest interaction is between the resistances of the CVS R and the ECMO speed and between the locations of the access and return lines of the CRRT circuit. The value of $S_2$ also indicates that higher order interactions contribute to $S_T$. There could be several reasons for this. Firstly, higher ECMO flows increase the effect of different locations for the access and return lines of the CRRT circuit, as we have already seen in the results. In addition, vascular resistance R may alter the flow dynamics so that blood is drawn and returned differently from the ECMO circuit. This would also suggest that the influence of the CRRT circuit and its possible combinations is different for each patient. In general, the significant interactions between CRRT access and return line locations with CVS resistances and ECMO speed demonstrate the complex interplay between combined lung and kidney support therapy and the patient's CVS.

In summary, our computational model provides results that are consistent with those reported in the literature and further emphasizes the complexity of ECMO and CRRT therapy. It describes hemodynamics on a patient-specific basis and can be adapted to routine clinical measurements. GSA-based parameter estimation has been shown to be fast and robust, and the credibility of the model has been demonstrated by applying it to a number of data points representing the full range of ECMO therapy conditions. Moreover, this demonstrated that the model can be easily recalibrated without the need for further user input, such as hyperparameter adjustment. This framework is open-source and promotes universal access to verified digital twins for other research questions, as advised in Viceconti et al. [34]. Parameter identification and model sensitivity which are core questions in the digital twin context, were at the focus of our investigations.

The implemented parameter identification pipeline yields a parameter set that reproduces the clinical measurements. However, this solution is not unique and cannot be guaranteed to truly reflect the specific patient condition. The use of the Markov chain Monte Carlo method, as proposed by Colunga et al. [48] and Argus et al. [59], could help to quantify the uncertainty of the derived parameter set. Although our implementation allows for just-in-time compilation using *JAX*, which significantly reduces computation time, it may be necessary for such an approach to create a surrogate of the LPM to provide clinicians with information quickly. The GSA enables to determine the most influential model parameters and to consider the interactions between them. However, it does not determine whether the effects of the input parameters on a specific model output are very similar or the same. This is referred to as parameter orthogonality and is discussed in detail in Colunga et al. [37] and Saxton et al. [42]. In a next step, an orthogonality analysis will be added to the GSA to ensure complete identifiability of the input parameters used for personalization.

Using this computational model, combined lung and kidney support therapy could now be tailored to a patient's individual condition. By applying the model to more patient data, more general conclusions could be drawn, which may ultimately lead to standardized guidelines. In addition to optimizing CRRT connection schemes, novel concepts such as the RenOx device, which integrates an oxygenator and a dialyzer in one system, would be of interest [60, 61]. Such an approach would drastically reduce the complexity highlighted in this study.

## 5. Conclusion

In this study, we were able to show that GSA-informed parameter identification can be used to reduce computational costs while ensuring a high-quality fit to V-A ECMO patient data. The developed cardiovascular model can be used to model many different conditions of patients receiving V-A ECMO. It was shown that the CRRT connection scheme has an influence on the hemodynamics of the patient and that the best choice is highly dependent on the operating point and setup of the ECMO system. By integrating a discrete distribution for the locations of both the access and return lines of the CRRT circuit into the sample of a GSA, we were able to systematically quantify their influence and compare their significance to both the patient's cardiovascular properties and the ECMO pump flow.

In summary, the results of this study indicate that the direct effect of CRRT access and return line locations within the ECMO circuit alone is minimal, but their combined effect when interacting with other parameters such as vascular resistance or ECMO pump speed is significant.

**Data availability**
The model and code to perform the GSA and parameter identification can be found at the GitHub link https://github.com/nikithiel/ECLIPSE.

**Acknowledgements**

This work was supported by the German Research Foundation (DFG) (project number 447746988, part of SPP 2014). Simulations were performed with computing resources granted by RWTH Aachen University under project rwth1463.

## 6. Supplements

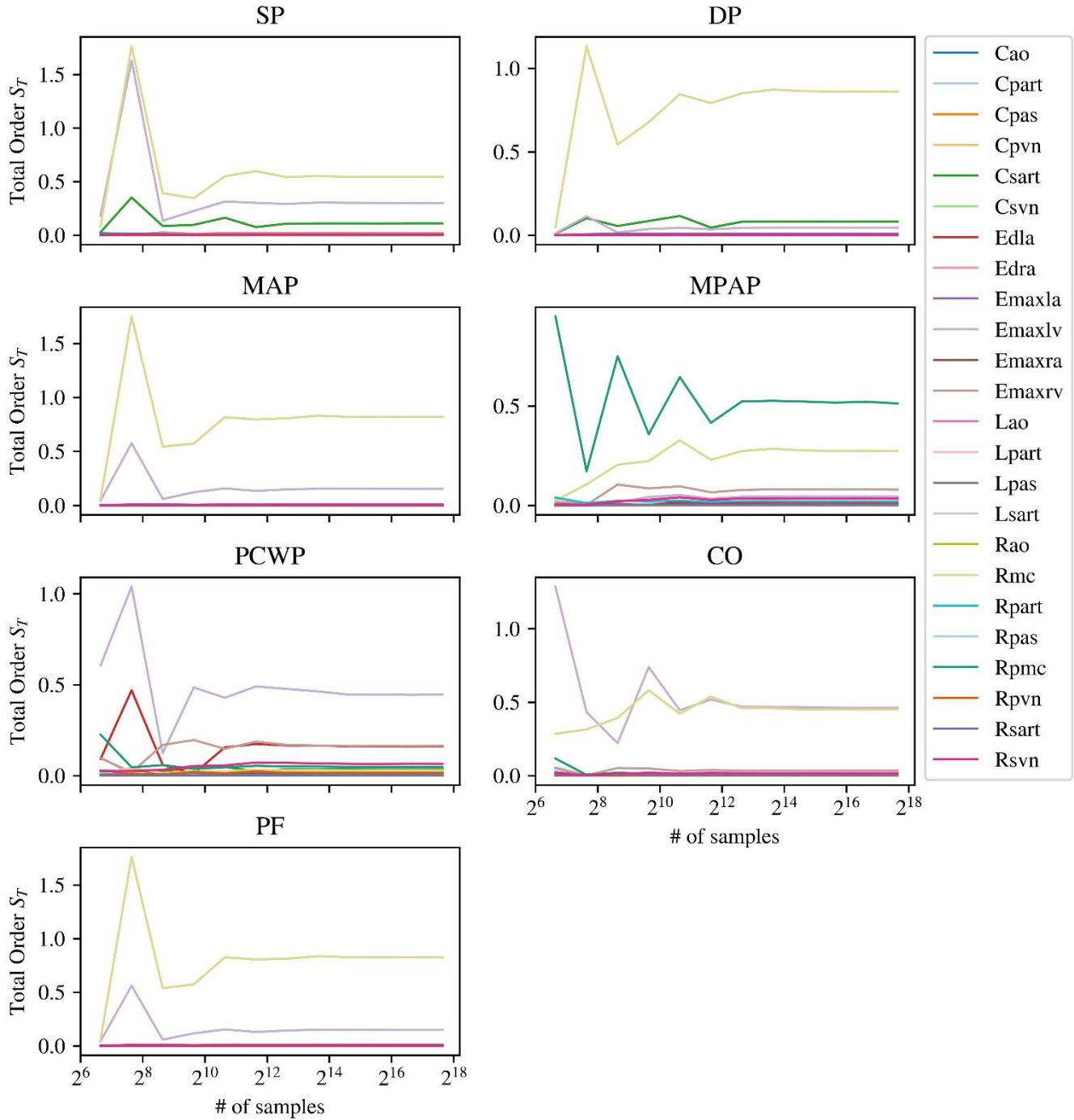

**Figure S-1:** Total order sensitivity indices for all clinical outputs and model parameters using different samples sizes. Cao: compliance of aorta, Cpart: compliance of pulmonary artery, Cpas: compliance of pulmonary aortic sinus, Cpvn: compliance of pulmonary vein, Csart: compliance of systemic artery, Csvn: compliance of systemic vein, Edla: diastolic elastance of left atrium, Edra: diastolic elastance of right atrium, Emaxla: systolic elastance of left atrium, Emaxlv: systolic elastance of left ventricle, Emaxra: systolic elastance of right atrium, Emaxrv: systolic elastance of right ventricle, Lao: inertance of aorta, Lpart: inertance of pulmonary artery, Lpas: inertance of pulmonary aortic sinus, Lsart: inertance of systemic artery, Rao: resistance of aorta, Rmc: resistance of systemic microcirculation, Rpart: resistance of pulmonary artery, Rpas: resistance of pulmonary aortic sinus, Rpmc: resistance of pulmonary microcirculation, Rpvn: resistance of pulmonary vein, Rsart: resistance of systemic artery, Rsvn: resistance of systemic vein, SP: systolic pressure, DP: diastolic pressure, MAP: mean arterial pressure, MPAP: mean pulmonary artery pressure, PCWP: pulmonary capillary wedge pressure, CO: cardiac output, PF: pump flow.

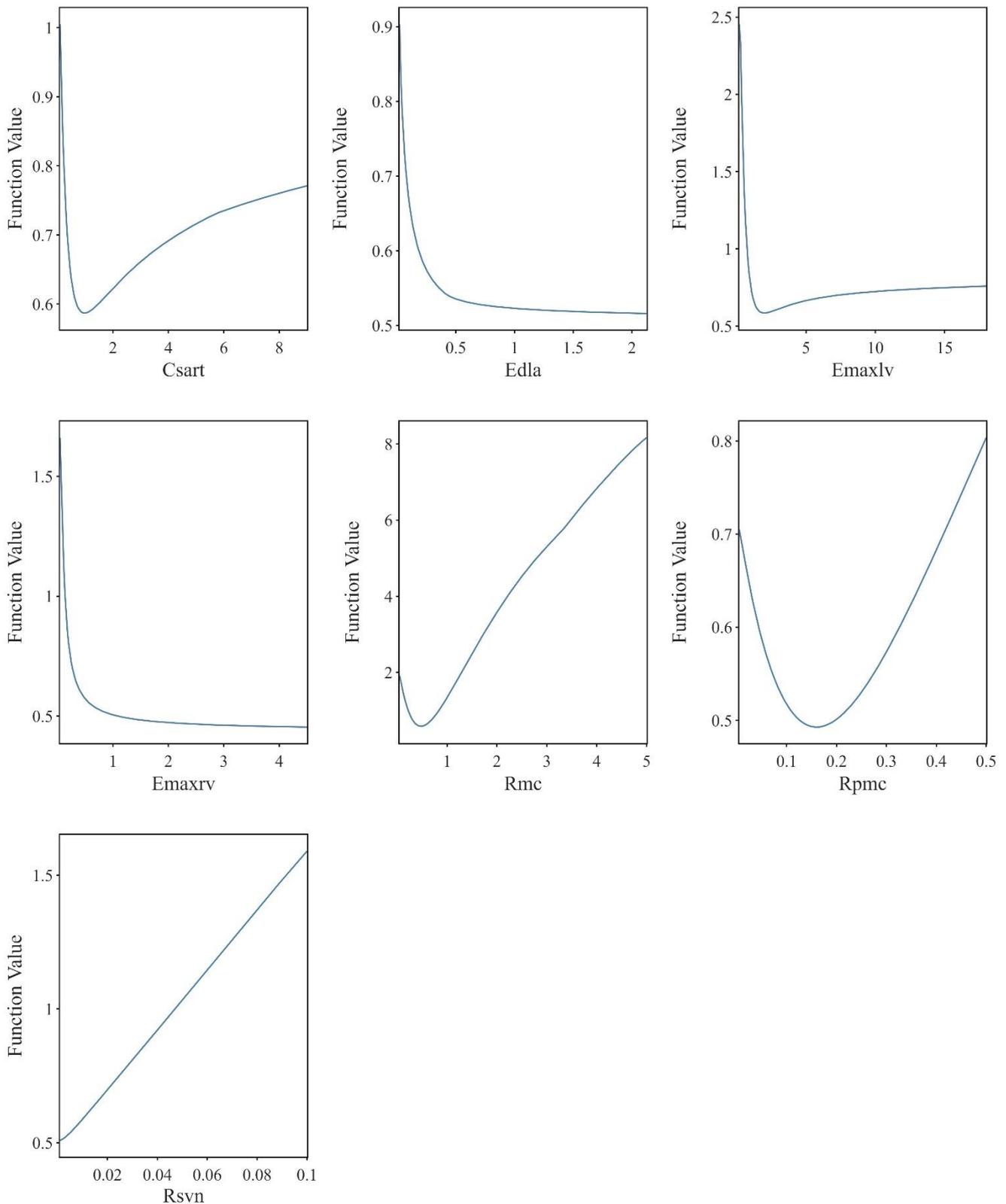

**Figure S-2:** Slice plots showing differentiability of objective function w.r.t parameters identified through Global sensitivity analysis and used for model calibration. Plot shows bounds of parameters. Csart: compliance of systemic artery, Edla: diastolic elastance of left atrium, Emaxlv: systolic elastance of left ventricle, Emaxrv: systolic elastance of right ventricle, Rmc: resistance of systemic microcirculation, Rpmc: resistance of pulmonary microcirculation, Rsvn: resistance of systemic vein.

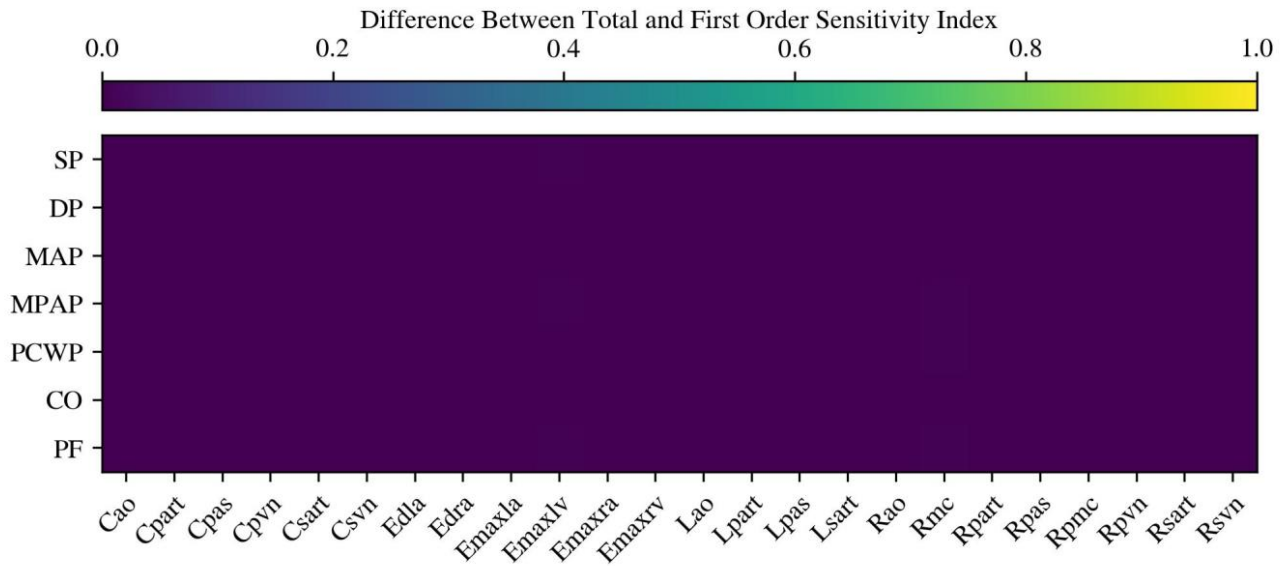

**Figure S-3:** Difference between Total and First order sensitivity indices for V-A ECMO patient with normal systolic pressure. Cao: compliance of aorta, Cpart: compliance of pulmonary artery, Cpas: compliance of pulmonary aortic sinus, Cpvn: compliance of pulmonary vein, Csart: compliance of systemic artery, Csvn: compliance of systemic vein, Edla: diastolic elastance of left atrium, Edra: diastolic elastance of right atrium, Emaxla: systolic elastance of left atrium, Emaxlv: systolic elastance of left ventricle, Emaxra: systolic elastance of right atrium, Emaxrv: systolic elastance of right ventricle, Lao: inertance of aorta, Lpart: inertance of pulmonary artery, Lpas: inertance of pulmonary aortic sinus, Lsart: inertance of systemic artery, Rao: resistance of aorta, Rmc: resistance of systemic microcirculation, Rpart: resistance of pulmonary artery, Rpas: resistance of pulmonary aortic sinus, Rpmc: resistance of pulmonary microcirculation, Rpvn: resistance of pulmonary vein, Rsart: resistance of systemic artery, Rsvn: resistance of systemic vein, SP: systolic pressure, DP: diastolic pressure, MAP: mean arterial pressure, MPAP: mean pulmonary artery pressure, PCWP: pulmonary capillary wedge pressure, CO: cardiac output, PF: pump flow.

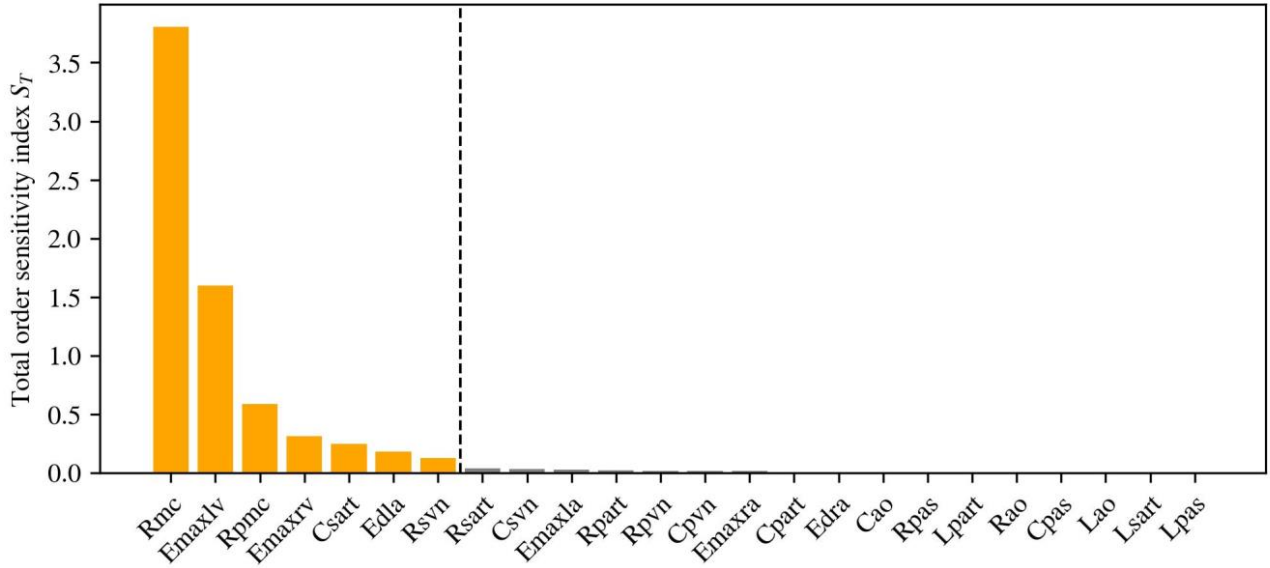

**Figure S-4:** Cumulative Total order sensitivity indices for each model parameter for patient with normal systolic pressure. Showing the selected parameters for model calibration. Cao: compliance of aorta, Cpart: compliance of pulmonary artery, Cpas: compliance of pulmonary aortic sinus, Cpvn: compliance of pulmonary vein, Csart: compliance of systemic artery, Csvn: compliance of systemic vein, Edla: diastolic elastance of left atrium, Edra: diastolic elastance of right atrium, Emaxla: systolic elastance of left atrium, Emaxlv: systolic elastance of left ventricle, Emaxra: systolic elastance of right atrium, Emaxrv: systolic elastance of right ventricle, Lao: inertance of aorta, Lpart: inertance of pulmonary artery, Lpas: inertance of pulmonary aortic sinus, Lsart: inertance of systemic artery, Rao: resistance of aorta, Rmc: resistance of systemic microcirculation, Rpart: resistance of pulmonary artery, Rpas: resistance of pulmonary aortic sinus, Rpmc: resistance of pulmonary microcirculation, Rpvn: resistance of pulmonary vein, Rsart: resistance of systemic artery, Rsvn: resistance of systemic vein, SP: systolic pressure, DP: diastolic pressure, MAP: mean arterial pressure, MPAP: mean pulmonary artery pressure, PCWP: pulmonary capillary wedge pressure, CO: cardiac output, PF: pump flow.

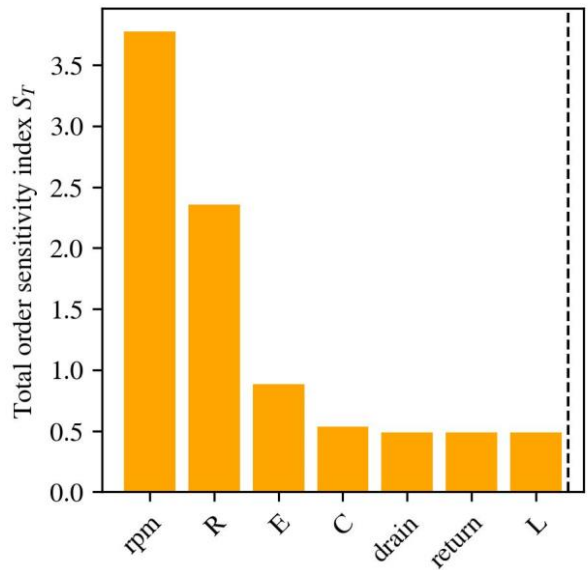

**Figure S-5:** Cumulative total order sensitivity indices for each model parameter for V-A ECMO patient from Global sensitivity analysis including CRRT connection scheme and ECMO pump speed (rpm). Grouping of model parameters into resistances (R), compliances (C), and inertances (L) of the cardiovascular system and properties of both left and right ventricle (E). SP: systolic pressure, DP: diastolic pressure, MAP: mean arterial pressure, MPAP: mean pulmonary artery pressure, PCWP: pulmonary capillary wedge pressure, CO: cardiac output, PF: pump flow.

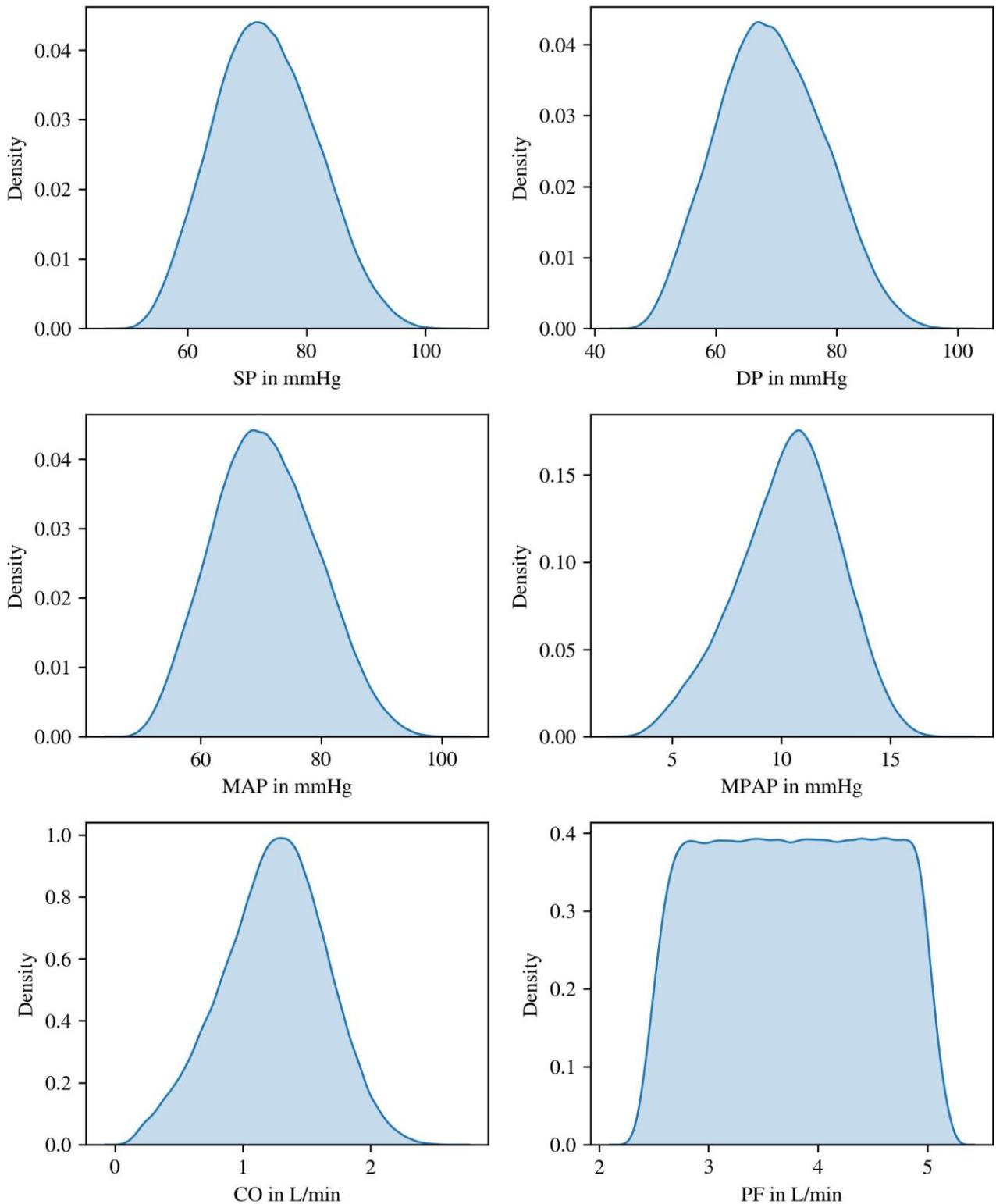

**Figure S-6:** Distribution of model outputs for sample used for Global sensitivity analysis including CRRT connection scheme and ECMO pump speed applied to V-A ECMO patient. SP: systolic pressure, DP: diastolic pressure, MAP: mean arterial pressure, MPAP: mean pulmonary artery pressure, PCWP: pulmonary capillary wedge pressure, CO: cardiac output, PF: pump flow.

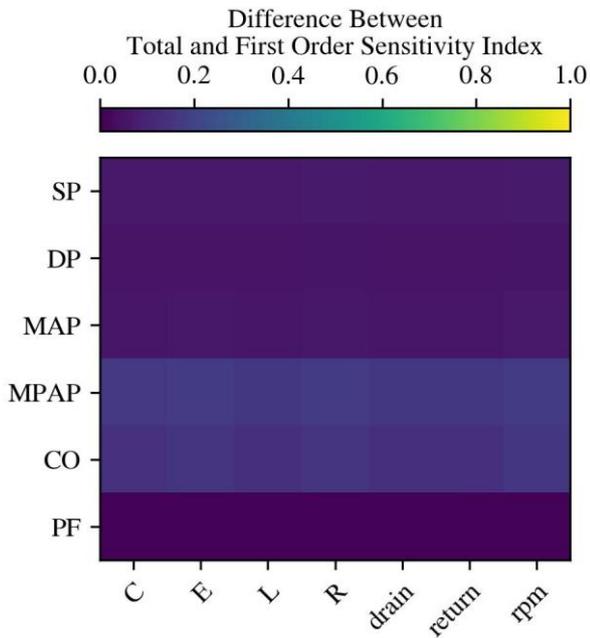

**Figure S-7:** Difference between Total and First order sensitivity indices for V-A ECMO patient from GSA including CRRT connection scheme and ECMO pump speed (rpm). Grouping of model parameters into resistances (R), compliances (C), and inertances (L) of the cardiovascular system and properties of both left and right ventricle (E). SP: systolic pressure, DP: diastolic pressure, MAP: mean arterial pressure, MPAP: mean pulmonary artery pressure, PCWP: pulmonary capillary wedge pressure, CO: cardiac output, PF: pump flow.

**Table S-1**
Model parameters used for the cardiovascular system.

| Parameter | Value | Description | Unit |
| --- | --- | --- | --- |
| ηB | 0.003 | Dynamic viscosity of blood | Pa s |
| ρB | 0.00106 | Density of blood | kg / mL |
| Rao | 0.003 | Resistance of aorta | (mmHg s) / mL |
| Rsart | 0.05 | Resistance of systemic artery | (mmHg s) / mL |
| Rmc | 0.5 | Resistance of systemic microcirculation | (mmHg s) / mL |
| Rsvn | 0.01 | Resistance of systemic vein | (mmHg s) / mL |
| Rpas | 0.002 | Resistance of pulmonary aortic sinus | (mmHg s) / mL |
| Rpart | 0.01 | Resistance of pulmonary artery | (mmHg s) / mL |
| Rpmc | 0.05 | Resistance of pulmonary microcirculation | (mmHg s) / mL |
| Rpvn | 0.006 | Resistance of pulmonary vein | (mmHg s) / mL |
| Cao | 0.08 | Compliance of aorta | mL / mmHg |
| Csart | 0.9 | Compliance of systemic artery | mL / mmHg |
| Csvn | 20.5 | Compliance of systemic vein | mL / mmHg |
| Cpas | 0.18 | Compliance of pulmonary aortic sinus | mL / mmHg |
| Cpart | 3.8 | Compliance of pulmonary artery | mL / mmHg |
| Cpvn | 20.5 | Compliance of pulmonary vein | mL / mmHg |
| Lao | 0.000062 | Inertance of aorta | mmHg s^2 / mL |
| Lsart | 0.0017 | Inertance of systemic artery | mmHg s^2 / mL |
| Lpas | 0.000052 | Inertance of pulmonary aortic sinus | mmHg s^2 / mL |
| Lpart | 0.0017 | Inertance of pulmonary artery | mmHg s^2 / mL |
| CQtri | 400 | Flow coefficient tricuspid valve | mL / (s mmHg^1/2) |
| CQpa | 350 | Flow coefficient pulmonary valve | mL / (s mmHg^1/2) |
| CQmi | 400 | Flow coefficient mitral valve | mL / (s mmHg^1/2) |
| CQao | 350 | Flow coefficient aortic valve | mL / (s mmHg^1/2) |
| Emaxra | 0.225 | Systolic elastance of right atrium | mmHg / mL |
| Edra | 0.1575 | Diastolic elastance of right atrium | mmHg / mL |
| Emaxrv | 0.45 | Systolic elastance of right ventricle | mmHg / mL |
| RV_Pd_beta | 2.5 | Right ventricle pressure-volume curve coefficient | mmHg |
| RV_Pd_kappa | 0.028 | Right ventricle pressure-volume curve coefficient | 1 / mL |
| RV_Pd_alpha | 0.047 | Right ventricle pressure-volume curve coefficient | mmHg |
| Emaxla | 0.45 | Systolic elastance of left atrium | mmHg / mL |
| Edla | 0.2125 | Diastolic elastance of left atrium | mmHg / mL |
| Emaxlv | 1.8 | Systolic elastance of left ventricle | mmHg / mL |
| LV_Pd_beta | 2.5 | Left ventricle pressure-volume curve coefficient | mmHg |
| LV_Pd_kappa | 0.033 | Left ventricle pressure-volume curve coefficient | 1 / mL |
| LV_Pd_alpha | 0.064 | Left ventricle pressure-volume curve coefficient | mmHg |

**Table S-2**
Model parameters used for the ECMO circuit.

| Parameter | Value | Description | Unit |
|---|---|---|---|
| Decmotudp | 0.9625 | Diameter (3/8") of tubing drainage - pump | cm |
| Lecmotudp | 200 | Length of tubing drainage - pump | cm |
| Decmotupo | 0.9625 | Diameter (3/8") of tubing pump - oxygenator | cm |
| Lecmotupo | 20 | Length of tubing pump - oxygenator | cm |
| Decmotuor | 0.9625 | Diameter (3/8") of tubing oxygenator - return | cm |
| Lecmotuor | 200 | Length of tubing oxygenator - return | cm |
| Recmooxy | 0.6744 | Resistance of oxygenator | (mmHg s) / mL |
| Cecmodrain | 0.1 | Compliance of drainage cannula | mL / mmHg |
| Cecmotudp | 0.1 | Compliance of tubing drainage - pump | mL / mmHg |
| Cecmotupo | 0.1 | Compliance of tubing pump - oxygenator | mL / mmHg |
| Cecmooxy | 0.008 | Compliance of oxygenator | mL / mmHg |
| Cecmotuor | 0.1 | Compliance of tubing oxygenator - return | mL / mmHg |
| Cecmoreturn | 0.1 | Compliance of return cannula | mL / mmHg |

**Table S-3**
Model parameters used for the CRRT circuit.

| Parameter | Value | Description | Unit |
|---|---|---|---|
| Lcrrttuin | 250 | Length of access line | cm |
| Dcrrttuin | 0.3175 | Diameter (1/8") of access line | cm |
| Lcrrttupf | 100 | Length of tubing pump - dialyzer | cm |
| Dcrrttupf | 0.3175 | Diameter (1/8") of tubing pump - dialyzer | cm |
| Lcrrttuout | 250 | Length of return line | cm |
| Dcrrttuout | 0.3175 | Diameter (1/8") of return line | cm |
| Rcrrtfil | 15.75 | Resistance of dialyzer | (mmHg s) / mL |
| Ccrrttuin | 0.1 | Compliance of access line | mL / mmHg |
| Ccrrttupf | 0.1 | Compliance of tubing pump - dialyzer | mL / mmHg |
| Ccrrtfil | 0.008 | Compliance of dialyzer | mL / mmHg |
| Ccrrttuout | 0.1 | Compliance of return line | mL / mmHg |

**Table S-4**
Fitted model parameters of specific patient used for the analysis of different CRRT connection schemes.

| Parameter | Value | Description | Unit |
|---|---|---|---|
| Csart | 3.1852 | Compliance of systemic artery | mL / mmHg |
| Edla | 2.1167 | Diastolic elastance of left atrium | mmHg / mL |
| Emaxlv | 1.3297 | Systolic elastance of left ventricle | mmHg / mL |
| Emaxrv | 0.10791 | Systolic elastance of right ventricle | mmHg / mL |
| Rmc | 0.75989 | Resistance of systemic microcirculation | (mmHg s) / mL |
| Rpmc | 0.3369 | Resistance of pulmonary microcirculation | (mmHg s) / mL |
| Rsvn | 0.00304 | Resistance of systemic vein | (mmHg s) / mL |

**Tables**

**Table 1**
Measured hemodynamic parameters for 30 data points from eight V-A ECMO patients.

| Parameter | Median | (Min, Max) |
|---|---|---|
| Systolic pressure (SP) | 111.0 | (70.0, 183.0) mmHg |
| Diastolic pressure (DP) | 69.5 | (42.0, 88.0) mmHg |
| Mean arterial pressure (MAP) | 79.0 | (55.0, 113.0) mmHg |
| Mean pulmonary artery pressure (MPAP) | 19.5 | (6.0, 30.0) mmHg |
| Pulmonary capillary wedge pressure (PCWP) | 11.0 | (7.0, 15.0) mmHg |
| Cardiac output (CO) | 3.4 | (1.5, 7.5) L/min |
| Pump flow (PF) | 2.5 | (1.2, 3.8) mmHg |

**Table 2**
Cumulative Total order sensitivity indices [$S_T$] for most influential model parameters for patient with normal systolic pressure. Rmc: resistance of the systemic microcirculation, Emaxlv: systolic elastance of the left ventricle, Rpmc: resistance of the pulmonary microcirculation, Emaxrv: systolic elastance of the right ventricle, Csart: compliance of the systemic artery, Edla: diastolic elastance of the left atrium, Rsvn: resistance of the systemic vein.

| Parameter | $\sum S_T$ |
|---|---|
| Rmc | 3.80 |
| Emaxlv | 1.60 |
| Rpmc | 0.60 |
| Emaxrv | 0.32 |
| Csart | 0.25 |
| Edla | 0.19 |
| Rsvn | 0.14 |

**Table S-1**
Model parameters used for the cardiovascular system.

| Parameter | Value | Description | Unit |
|---|---|---|---|
| ηB | 0.003 | Dynamic viscosity of blood | Pa s |
| ρB | 0.00106 | Density of blood | kg / mL |
| Rao | 0.003 | Resistance of aorta | (mmHg s) / mL |
| Rsart | 0.05 | Resistance of systemic artery | (mmHg s) / mL |
| Rmc | 0.5 | Resistance of systemic microcirculation | (mmHg s) / mL |
| Rsvn | 0.01 | Resistance of systemic vein | (mmHg s) / mL |
| Rpas | 0.002 | Resistance of pulmonary aortic sinus | (mmHg s) / mL |
| Rpart | 0.01 | Resistance of pulmonary artery | (mmHg s) / mL |
| Rpmc | 0.05 | Resistance of pulmonary microcirculation | (mmHg s) / mL |
| Rpvn | 0.006 | Resistance of pulmonary vein | (mmHg s) / mL |
| Cao | 0.08 | Compliance of aorta | mL / mmHg |
| Csart | 0.9 | Compliance of systemic artery | mL / mmHg |
| Csvn | 20.5 | Compliance of systemic vein | mL / mmHg |
| Cpas | 0.18 | Compliance of pulmonary aortic sinus | mL / mmHg |
| Cpart | 3.8 | Compliance of pulmonary artery | mL / mmHg |
| Cpvn | 20.5 | Compliance of pulmonary vein | mL / mmHg |
| Lao | 0.000062 | Inertance of aorta | mmHg s^2 / mL |
| Lsart | 0.0017 | Inertance of systemic artery | mmHg s^2 / mL |
| Lpas | 0.000052 | Inertance of pulmonary aortic sinus | mmHg s^2 / mL |
| Lpart | 0.0017 | Inertance of pulmonary artery | mmHg s^2 / mL |
| CQtri | 400.0 | Flow coefficient tricuspid valve | mL / (s mmHg^1/2) |
| CQpa | 350.0 | Flow coefficient pulmonary valve | mL / (s mmHg^1/2) |
| CQmi | 400.0 | Flow coefficient mitral valve | mL / (s mmHg^1/2) |
| CQao | 350.0 | Flow coefficient aortic valve | mL / (s mmHg^1/2) |
| Emaxra | 0.225 | Systolic elastance of right atrium | mmHg / mL |
| Edra | 0.1575 | Diastolic elastance of right atrium | mmHg / mL |
| Emaxrv | 0.45 | Systolic elastance of right ventricle | mmHg / mL |
| RV_Pd_beta | 2.5 | Right ventricle pressure-volume curve coefficient | mmHg |
| RV_Pd_kappa | 0.028 | Right ventricle pressure-volume curve coefficient | 1 / mL |
| RV_Pd_alpha | 0.047 | Right ventricle pressure-volume curve coefficient | mmHg |
| Emaxla | 0.45 | Systolic elastance of left atrium | mmHg / mL |
| Edla | 0.2125 | Diastolic elastance of left atrium | mmHg / mL |
| Emaxlv | 1.8 | Systolic elastance of left ventricle | mmHg / mL |
| LV_Pd_beta | 2.5 | Left ventricle pressure-volume curve coefficient | mmHg |
| LV_Pd_kappa | 0.033 | Left ventricle pressure-volume curve coefficient | 1 / mL |
| LV_Pd_alpha | 0.064 | Left ventricle pressure-volume curve coefficient | mmHg |

**Table S-2**
Model parameters used for the ECMO circuit.

| Parameter | Value | Description | Unit |
|---|---|---|---|
| Decmotudp | 0.9625 | Diameter (3/8") of tubing drainage - pump | cm |
| Lecmotudp | 200 | Length of tubing drainage - pump | cm |
| Decmotupo | 0.9625 | Diameter (3/8") of tubing pump - oxygenator | cm |
| Lecmotupo | 20 | Length of tubing pump - oxygenator | cm |
| Decmotuor | 0.9625 | Diameter (3/8") of tubing oxygenator - return | cm |
| Lecmotuor | 200 | Length of tubing oxygenator - return | cm |
| Recmooxy | 0.6744 | Resistance of oxygenator | (mmHg s) / mL |
| Cecmodrain | 0.1 | Compliance of drainage cannula | mL / mmHg |
| Cecmotudp | 0.1 | Compliance of tubing drainage - pump | mL / mmHg |
| Cecmotupo | 0.1 | Compliance of tubing pump - oxygenator | mL / mmHg |
| Cecmooxy | 0.008 | Compliance of oxygenator | mL / mmHg |
| Cecmotuor | 0.1 | Compliance of tubing oxygenator - return | mL / mmHg |
| Cecmoreturn | 0.1 | Compliance of return cannula | mL / mmHg |

**Table S-3**
Model parameters used for the CRRT circuit.

| Parameter | Value | Description | Unit |
|---|---|---|---|
| Lcrrttuin | 250 | Length of access line | cm |
| Dcrrttuin | 0.3175 | Diameter (1/8") of access line | cm |
| Lcrrttupf | 100 | Length of tubing pump - dialyzer | cm |
| Dcrrttupf | 0.3175 | Diameter (1/8") of tubing pump - dialyzer | cm |
| Lcrrttuout | 250 | Length of return line | cm |
| Dcrrttuout | 0.3175 | Diameter (1/8") of return line | cm |
| Rcrrtfil | 15.75 | Resistance of dialyzer | (mmHg s) / mL |
| Ccrrttuin | 0.1 | Compliance of access line | mL / mmHg |
| Ccrrttupf | 0.1 | Compliance of tubing pump - dialyzer | mL / mmHg |
| Ccrrtfil | 0.008 | Compliance of dialyzer | mL / mmHg |
| Ccrrttuout | 0.1 | Compliance of return line | mL / mmHg |

**Table S-4**
Fitted model parameters of specific patient used for the analysis of different CRRT connection schemes.

| Parameter | Value | Description | Unit |
|---|---|---|---|
| Csart | 3.1852 | Compliance of systemic artery | mL / mmHg |
| Edla | 2.1167 | Diastolic elastance of left atrium | mmHg / mL |
| Emaxlv | 1.3297 | Systolic elastance of left ventricle | mmHg / mL |
| Emaxrv | 0.10791 | Systolic elastance of right ventricle | mmHg / mL |
| Rmc | 0.75989 | Resistance of systemic microcirculation | (mmHg s) / mL |
| Rpmc | 0.3369 | Resistance of pulmonary microcirculation | (mmHg s) / mL |
| Rsvn | 0.00304 | Resistance of systemic vein | (mmHg s) / mL |

**Figure captions**

**Figure 1:** Overview of the lumped parameter model. Cardiovascular system inspired by Shi et al. [1]. pas: pulmonary aortic sinus, part: pulmonary artery, pmc: pulmonary microcirculation, pvn: pulmonary vein, ao: aorta, sart: systemic artery, smc: systemic microcirculation, svn: systemic vein, TV: tricuspid valve, PV: pulmonary valve, MV: mitral valve, AV: aortic valve, ECMO: extracorporeal membrane oxygenation, CRRT: continuous renal replacement therapy.

**Figure 2:** Variations of CRRT connections to the ECMO circuit in CRRT-flow direction: a) post pump – pre oxygenator (pre oxy), b) pre oxy – pre pump, c) post oxy – pre pump, d) post oxy – pre return cannula (pre RC), e) pre oxy – pre RC, f) pre oxy – post oxy, g) post oxy – pre oxy, h) pre pump – post drainage cannula (post DC), i) post pump – pre pump.
Of note: position of tubing e.g. between pump and oxygenator (compare (b) and (i)) is also taken into account by making a distinction between post pump and pre oxy.

**Figure 3:** Total order sensitivity indices for patient with normal systolic pressure. Cao: compliance of aorta, Cpart: compliance of pulmonary artery, Cpas: compliance of pulmonary aortic sinus, Cpvn: compliance of pulmonary vein, Csart: compliance of systemic artery, Csvn: compliance of systemic vein, Edla: diastolic elastance of left atrium, Edra: diastolic elastance of right atrium, Emaxla: systolic elastance of left atrium, Emaxlv: systolic elastance of left ventricle, Emaxra: systolic elastance of right atrium, Emaxrv: systolic elastance of right ventricle, Lao: inertance of aorta, Lpart: inertance of pulmonary artery, Lpas: inertance of pulmonary aortic sinus, Lsart: inertance of systemic artery, Rao: resistance of aorta, Rmc: resistance of systemic microcirculation, Rpart: resistance of pulmonary artery, Rpas: resistance of pulmonary aortic sinus, Rpmc: resistance of pulmonary microcirculation, Rpvn: resistance of pulmonary vein, Rsart: resistance of systemic artery, Rsvn: resistance of systemic vein, SP: systolic pressure, DP: diastolic pressure, MAP: mean arterial pressure, MPAP: mean pulmonary artery pressure, PCWP: pulmonary capillary wedge pressure, CO: cardiac output, PF: pump flow.

**Figure 4:** a) + b) Scatter plot of simulated data against measurement data for n = 30 datapoints of 8 V-A ECMO patients. c) + d) Box plots for simulated and measurement data for n = 30 datapoints of 8 V-A ECMO patients. SP: systolic pressure, DP: diastolic pressure, MAP: mean arterial pressure, MPAP: mean pulmonary artery pressure, PCWP: pulmonary capillary wedge pressure, CO: cardiac output, PF: pump flow.

**Figure 5:** Influence of different CRRT connections schemes on a) cardiovascular system, b) right ventricular (RV) PV-loop and pressure of both access and return line of the CRRT circuit in c) and d), respectively. Common pressure alarms of CRRT circuit displayed in black dashed lines. Orange lines represent the configuration pre pump – post DC, blue lines post pump – pre oxy, grey lines the other configurations. DC: drainage cannula, RC: return cannula.

**Figure 6:** Influence of different CRRT connections schemes with increasing ECMO pump flow on a) cardiovascular system, b) pressure in ECMO drainage cannula (DC) and both access and return line of the CRRT circuit in c) and d), respectively. Common pressure alarms of CRRT circuit displayed in black dashed lines. Orange lines represent the configuration pre pump – post DC, blue lines post pump – pre oxy, grey lines the other configurations.

**Figure 7:** Total order sensitivity indices for V-A ECMO patient from Global sensitivity analysis including CRRT connection scheme and ECMO pump speed. Grouping of model parameters into resistances (R), compliances (C), and inertances (L) of the cardiovascular system and properties of both left and right ventricle (E).

**Figure S-1:** Total order sensitivity indices for all clinical outputs and model parameters using different samples sizes. Cao: compliance of aorta, Cpart: compliance of pulmonary artery, Cpas: compliance of pulmonary aortic sinus, Cpvn: compliance of pulmonary vein, Csart: compliance of systemic artery, Csvn: compliance of systemic vein, Edla: diastolic elastance of left atrium, Edra: diastolic elastance of right atrium, Emaxla: systolic elastance of left atrium, Emaxlv: systolic elastance of left ventricle, Emaxra: systolic elastance of right atrium, Emaxrv: systolic elastance of right ventricle, Lao: inertance of aorta, Lpart: inertance of pulmonary artery, Lpas: inertance of pulmonary aortic sinus, Lsart: inertance of systemic artery, Rao: resistance of aorta, Rmc: resistance of systemic microcirculation, Rpart: resistance of pulmonary artery, Rpas: resistance of pulmonary aortic sinus, Rpmc: resistance of pulmonary microcirculation, Rpvn: resistance of pulmonary vein, Rsart: resistance of systemic artery, Rsvn: resistance of systemic

vein, SP: systolic pressure, DP: diastolic pressure, MAP: mean arterial pressure, MPAP: mean pulmonary artery pressure, PCWP: pulmonary capillary wedge pressure, CO: cardiac output, PF: pump flow.

**Figure S-2:** Slice plots showing differentiability of objective function w.r.t parameters identified through Global sensitivity analysis and used for model calibration. Plot shows bounds of parameters. Csart: compliance of systemic artery, Edla: diastolic elastance of left atrium, Emaxlv: systolic elastance of left ventricle, Emaxrv: systolic elastance of right ventricle, Rmc: resistance of systemic microcirculation, Rpmc: resistance of pulmonary microcirculation, Rsvn: resistance of systemic vein.

**Figure S-3:** Difference between Total and First order sensitivity indices for V-A ECMO patient with normal systolic pressure. Cao: compliance of aorta, Cpart: compliance of pulmonary artery, Cpas: compliance of pulmonary aortic sinus, Cpvn: compliance of pulmonary vein, Csart: compliance of systemic artery, Csvn: compliance of systemic vein, Edla: diastolic elastance of left atrium, Edra: diastolic elastance of right atrium, Emaxla: systolic elastance of left atrium, Emaxlv: systolic elastance of left ventricle, Emaxra: systolic elastance of right atrium, Emaxrv: systolic elastance of right ventricle, Lao: inertance of aorta, Lpart: inertance of pulmonary artery, Lpas: inertance of pulmonary aortic sinus, Lsart: inertance of systemic artery, Rao: resistance of aorta, Rmc: resistance of systemic microcirculation, Rpart: resistance of pulmonary artery, Rpas: resistance of pulmonary aortic sinus, Rpmc: resistance of pulmonary microcirculation, Rpvn: resistance of pulmonary vein, Rsart: resistance of systemic artery, Rsvn: resistance of systemic vein, SP: systolic pressure, DP: diastolic pressure, MAP: mean arterial pressure, MPAP: mean pulmonary artery pressure, PCWP: pulmonary capillary wedge pressure, CO: cardiac output, PF: pump flow.

**Figure S-4:** Cumulative Total order sensitivity indices for each model parameter for patient with normal systolic pressure. Showing the selected parameters for model calibration. Cao: compliance of aorta, Cpart: compliance of pulmonary artery, Cpas: compliance of pulmonary aortic sinus, Cpvn: compliance of pulmonary vein, Csart: compliance of systemic artery, Csvn: compliance of systemic vein, Edla: diastolic elastance of left atrium, Edra: diastolic elastance of right atrium, Emaxla: systolic elastance of left atrium, Emaxlv: systolic elastance of left ventricle, Emaxra: systolic elastance of right atrium, Emaxrv: systolic elastance of right ventricle, Lao: inertance of aorta, Lpart: inertance of pulmonary artery, Lpas: inertance of pulmonary aortic sinus, Lsart: inertance of systemic artery, Rao: resistance of aorta, Rmc: resistance of systemic microcirculation, Rpart: resistance of pulmonary artery, Rpas: resistance of pulmonary aortic sinus, Rpmc: resistance of pulmonary microcirculation, Rpvn: resistance of pulmonary vein, Rsart: resistance of systemic artery, Rsvn: resistance of systemic vein, SP: systolic pressure, DP: diastolic pressure, MAP: mean arterial pressure, MPAP: mean pulmonary artery pressure, PCWP: pulmonary capillary wedge pressure, CO: cardiac output, PF: pump flow.

**Figure S-5:** Cumulative total order sensitivity indices for each model parameter for V-A ECMO patient from Global sensitivity analysis including CRRT connection scheme and ECMO pump speed (rpm). Grouping of model parameters into resistances (R), compliances (C), and inertances (L) of the cardiovascular system and properties of both left and right ventricle (E). SP: systolic pressure, DP: diastolic pressure, MAP: mean arterial pressure, MPAP: mean pulmonary artery pressure, PCWP: pulmonary capillary wedge pressure, CO: cardiac output, PF: pump flow.

**Figure S-6:** Distribution of model outputs for sample used for Global sensitivity analysis including CRRT connection scheme and ECMO pump speed applied to V-A ECMO patient. SP: systolic pressure, DP: diastolic pressure, MAP: mean arterial pressure, MPAP: mean pulmonary artery pressure, PCWP: pulmonary capillary wedge pressure, CO: cardiac output, PF: pump flow.

**Figure S-7:** Difference between Total and First order sensitivity indices for V-A ECMO patient from GSA including CRRT connection scheme and ECMO pump speed (rpm). Grouping of model parameters into resistances (R), compliances (C), and inertances (L) of the cardiovascular system and properties of both left and right ventricle (E). SP: systolic pressure, DP: diastolic pressure, MAP: mean arterial pressure, MPAP: mean pulmonary artery pressure, PCWP: pulmonary capillary wedge pressure, CO: cardiac output, PF: pump flow.